\documentclass[journal]{IEEEtran}
\ifCLASSINFOpdf
\else
\fi
\usepackage{multirow}
\usepackage{amsthm}
\usepackage{graphicx}
\usepackage{color}
\usepackage{epstopdf}
\usepackage{amsfonts}
\usepackage[caption=false, font=footnotesize]{subfig}
\usepackage{cite}
\usepackage{hyperref}
\usepackage{amssymb}
\usepackage{amsmath}
\usepackage{algpseudocode,algorithm}
\usepackage{setspace}
\newtheorem{lemma}{Lemma}
\newtheorem{remark}{Remark}
\newtheorem{theorem}{Theorem}
\newtheorem{prop}{Proposition}
\newtheorem{corollary}{Corollary}
\newtheorem{example}{Example}
\DeclareMathSizes{10}{9}{7}{6}
\IEEEaftertitletext{\vspace{-2\baselineskip}}
\usepackage{tikz}
\definecolor{lime}{HTML}{A6CE39}
\DeclareRobustCommand{\orcidicon}{%
    \begin{tikzpicture}
    \draw[lime, fill=lime] (0,0) 
    circle [radius=0.16] 
    node[white] {{\fontfamily{qag}\selectfont \tiny ID}};    \draw[white, fill=white] (-0.0625,0.095) 
    circle [radius=0.007];    \end{tikzpicture}
    \hspace{-2mm}}
\foreach \x in {A, ..., Z}{%
    \expandafter\xdef\csname orcid\x\endcsname{\noexpand\href{https://orcid.org/\csname orcidauthor\x\endcsname}{\noexpand\orcidicon}}
    }

\begin{document}
\title{Primitive Rateless Codes}
\author{Mahyar~Shirvanimoghaddam\orcidA{},~\IEEEmembership{Senior~Member,~IEEE}

\thanks{M. Shirvanimoghaddam is with the School of Electrical and Information Engineering, The University of Sydney, NSW 2006, Australia (e-mail: mahyar.shirvanimoghaddam@sydney.edu.au).

This article was presented in part at the 2021 IEEE International Symposium on Information Theory (ISIT).

This paper has been accepted for publication by IEEE. DOI (identifier) 10.1109/TCOMM.2021.3096961. Copyright (c) 2017 IEEE. Personal use of this material is permitted.  However, permission to use this material for any other purposes must be obtained from the IEEE by sending a request to pubs-permissions@ieee.org.}}
\maketitle

\begin{abstract}
In this paper, we propose primitive rateless (PR) codes. A PR code is characterized by the message length and a primitive polynomial over $\mathbf{GF}(2)$, which can generate a potentially limitless number of coded symbols. We show that codewords of a PR code truncated at any arbitrary length can be represented as subsequences of a maximum-length sequence ($m$-sequence). We characterize the Hamming weight distribution of PR codes and their duals and show that for a properly chosen primitive polynomial, the Hamming weight distribution of the PR code can be well approximated by the truncated binomial distribution. We further find a lower bound on the minimum Hamming weight of PR codes and show that there always exists a PR code that can meet this bound for any desired codeword length. We provide a list of primitive polynomials for message lengths up to $40$ and show that the respective PR codes closely meet the Gilbert-Varshamov bound at various rates. Simulation results show that PR codes can achieve similar block error rates as their BCH counterparts at various signal-to-noise ratios (SNRs) and code rates. PR codes are rate-compatible and can generate as many coded symbols as required; thus, demonstrating a truly rateless performance.
\end{abstract}
\IEEEpeerreviewmaketitle

\section{Introduction}
\IEEEPARstart{R}{ate}-compatible (RC) error-control codes are a set of codes with the same dimension but various code lengths and accordingly rates, where all symbols of the higher-rate code are a subset of the lower rate code \cite{davida1972forward}. Combined with RC codes, hybrid automatic repeat request (HARQ) schemes have been used in wireless communication systems to match the code rate to the channel condition by retransmitting incremental redundancy (IR) to the receiver \cite{huang2020syndrome}. The rate-matching procedure is crucial to support various requirements and be able to adapt to varying channel conditions. This becomes even more important for modern wireless systems, like the fifth generation (5G) of mobile communications standard that has established a framework to include services with a diverse range of requirements, such as ultra-reliable and low-latency communications (URLLC) and massive machine-type communications (mMTC), in addition to the traditional enhanced mobile broadband (eMBB) \cite{shirvanimoghaddam2018short}.

The most common way to construct RC codes is to use puncturing. A good low-rate code, referred to as the mother-code, is first constructed, and then some of the coded symbols are discarded to construct higher-rate codes. This approach has been applied to almost all codes and in particular to algebraic codes \cite{davida1972forward,wicker1994type}, convolutional codes \cite{hagenauer1988rate,lee1994new}, Turbo codes \cite{liu2003punctured,rowitch2000performance}, low-density parity-check (LDPC) codes \cite{ha2004rate,el2009design}, and Polar codes \cite{hong2017capacity,niu2013beyond}. The performance of the resulting code depends mainly on the puncturing pattern \cite{wang2021some}. Finding the best puncturing pattern is usually nontrivial and carried out through computer search. Moreover, most puncturing techniques are optimized according to the set of information bits; therefore, these methods cannot be used to design a family of rate-compatible punctured codes for IR-HARQ, which requires the same information set that should be used for all punctured codes from a mother code in the family \cite{hong2017capacity}. Furthermore, if the rate of the mother code is too low, puncturing is not likely to yield good high-rate codes. For example, although Polar codes can achieve the capacity of any specific binary-input symmetric channel, the rate-compatible construction
via successive puncturing is generally not capacity-achieving \cite{hong2017capacity}.

Extending is another approach to construct RC codes \cite{krishna1987new}. A good high-rate code is first constructed, then parity-check symbols are successively added to generate lower-rate codes. The construction of lower rate codes is to find new codes with a good minimum Hamming weight. This approach has been used to construct RC-LDPC codes \cite{chen2015protograph,van2012design} and RC- Polar codes \cite{hong2017capacity,li2016capacity}. RC codes constructed using extending do not usually guarantee high minimum Hamming weight at lower rates, as the minimum weight at a particular rate depends on the original code.  Protograph-based Raptor-like LDPC (PBRL) codes were proposed in \cite{chen2015protograph}, which are a class of rate-compatible LDPC codes with extensive rate compatibility. The design of PBRL codes at short and long block lengths \cite{ranganathan2019quasi} have been based on optimizing the iterative decoding threshold of the protograph at various design rates. In particular, each additional parity bit in the protograph is explicitly designed to optimize the density evolution threshold. These codes have been standardized for 5G enhanced mobile broadband \cite{shirvanimoghaddam2018short} for the data channel.

A more general approach based on extending is to use rateless codes. Also known as Fountain codes, rateless codes can generate a potentially limitless number of coded symbols for a given set of input symbols. The coded symbols are usually produced independently and randomly. The receiver can then recover the original input symbols from any subset of received symbols given the length of the subset is sufficiently large. Luby transform (LT) codes \cite{luby2002lt} were the first practical realization of Fountain codes, where each output symbol is generated by adding $d$ randomly chosen input symbols, where $d$ is obtained from a predefined probability distribution function, referred to as the degree distribution.  LT codes suffer from error-floor, which is mainly due to random selection of input symbols when generating coded symbols. Rapid tornado (Raptor) codes \cite{shokrollahi2006raptor} solve this problem by adding a high-rate precoder, usually a LDPC code, to an LT code. When the degree distribution is chosen properly, the decoder can recover the original $k$ input symbols from any $n$ Raptor coded symbols as long as $n$ is slightly larger than $k$.  The encoding and decoding complexity of Raptor codes increases linearly with the message length \cite{shokrollahi2006raptor}. 

Raptor codes can be used over noisy channels \cite{palanki2004rateless,Shirvani2016Low}. Authors in \cite{etesami2006raptor} showed that unlike the erasure channel where a universal degree distribution can be optimized for all erasure rates, the optimized degree distributions of Raptor codes over BI-AWGN and binary symmetric channel (BSC) depend on the channel; therefore, non-universal. To address this problem, several approaches based on adaptive degree distribution design were proposed in the literature \cite{kuo2014design,jayasooriya2018design}. The design and analysis are, however, valid only for asymptotically long block lengths.  In binary rateless codes, such as Raptor and LT codes, each new coded symbol is generated through a random process. However, this does not guarantee that the minimum Hamming weight of the code is increased by adding a new parity symbol. It is also probable that a redundant coded symbol is being generated. In fact, the code weight spectrum or the minimum Hamming weight have not been the design criteria for these codes, as they were mainly designed for asymptotically long message sizes. 

It is now recognized that bit-level granularity of the codeword size and code operating rate is desired for 5G and beyond \cite{3gppchannelcode}. The actual coding rate used in transmission could not be restricted and optimized for specified ranges \cite{3gppchannelcode}. Designing and optimizing short block length codes have been recently attracted for being implemented on memory or power-constrained devices, mainly in the context of the Internet of Things applications and services. Existing RC codes are mainly constructed using puncturing and extending, which are shown to be sub-optimal for short block lengths.  Therefore, designing RC codes for short messages that support bit-level granularity of the codeword size and maintain a large minimum Hamming weight at various rates is still open. 

In this paper, we propose primitive rateless (PR) codes, which are mainly characterized by the message length $k$ and a primitive polynomial of degree $k$ over $\mathbf{GF}(2)$. We show that PR codes can be represented by using 1) a linear-feedback shift-register (LFSR) with connection polynomial $x^kp(1/x)$ and 2) Boolean functions. In fact, the codewords of a PR code are subsequences of a maximum-length sequence ($m$-sequence). We show that any two PR codes of dimension $k$ and truncated at length $n\ge 2k$, which are constructed by using two distinct primitive polynomials, do not have any non-zero codeword in common. We also characterize the average Hamming weight distribution of PR codes and develop a lower bound on the minimum Hamming weight which is very close to the Gilbert-Varshamov bound \cite{Jiang2004}. We  show that for any $k$, there exists at least one PR code that can meet this bound. We further find some good primitive polynomials for PR codes of dimension $k\le 40$, which can closely approach the Gilbert-Varshamov bound at various rates. Simulation results show that the PR code with a properly chosen primitive polynomial can achieve a similar block error rate (BLER) performance as the extended Bose, Chaudhuri, and Hocquenghem (eBCH) code counterpart. This is because while a PR codes has a lower minimum Hamming weight than the eBCH code, it has a lower number of low-weight codewords. Simulation results show that PR codes in a rateless setting can achieve a very high realized rate over a wide range of SNRs. PR codes can be designed for any message length and arbitrary rate and perform very close to finite block length bounds. They are rate-compatible and have a very simple encoding structure, unlike most rate-compatible codes designed based on puncturing a low-rate mother code, with mostly sub-optimal performance at various rates. 

The rest of the paper is organized as follows. Section II introduces primitive rateless (PR) codes and discusses some of their important properties. In Section III, we characterize the Hamming weight distribution of the dual of the PR code and then find the average Hamming weight distribution of PR codes. We also characterize the minimum Hamming weight of the PR codes. In Section IV, we explain how to choose the primitive polynomial for PR codes and provide a list of good primitive polynomials for message lengths up to $40$. Numerical results are presented in Section V. Finally, Section VI concludes the paper.

\section{Construction of the Primitive Rateless Code}
A primitive rateless (PR) code, denoted by $\mathrm{PR}(k,p(x))$, is characterized by the information block length $k$ and a binary primitive polynomial $p(x)=\sum_{i=0}^{k}p_{i}x^{i}$ of degree $k$, for $p_0=p_k=1$, which is the minimal polynomial of a primitive element $\alpha$ over $\mathbf{GF}(2^k)$. The generator matrix of a PR code $\mathrm{PR}(k,p(x))$ truncated at length $n$, which is denoted by $\mathrm{PR}(k,n,p(x))$, is constructed as follows:
\begin{align}
\mathbf{G}_n=\left[\alpha^0, \alpha^1, \cdots, \alpha^{n-1}\right],
\end{align}
where the $i^{th}$ column is the binary representation of $\alpha^{i-1}$, for $1\le i\le n$. Since $\alpha$ is a primitive element of $\mathbf{GF}(2^k)$, $\{0,1,\alpha,\alpha^2,\cdots,\alpha^{2^k-2}\}$ is the entire field $\mathbf{GF}(2^k)$ \cite{Lidl1997}.  The parity check matrix of $\mathrm{PR}(k,n,p(x))$ is given by:
\begin{equation}
\mathbf{H}_n=
\begin{bmatrix}
p_0&p_1&\cdots&p_{k}&0&0&\cdots\\
0&p_0&p_1&\cdots&p_{k}&0&\cdots\\
&\ddots&&\ddots&&\ddots&\\
0&\cdots&0&p_0&p_1&\cdots&p_{k}
\end{bmatrix},
\label{pmat}
\end{equation}
where the $i^{th}$ row is the $(i-1)^{th}$-order cyclic shift of the first row, for $2\le i\le n-k$. Since $\alpha$ is a primitive element and $\sum_{i=0}^{k-1}p_{i}\alpha^{i}=\alpha^{k}$, it can be easily verified that $\mathbf{G}_n\mathbf{H}_n^{\top}=\mathbf{0}$, where$~^\top$ is the matrix transpose operand. It is important to note that since $\alpha$ is a primitive $(2^k-1)$-root of unity in $\mathbf{GF}(2^k)$, i.e., $\alpha^{2^k-1}=1$, the columns of the generator matrix will be repeating for $n\ge 2^k-1$.

\begin{remark}[LFSR-based construction of PR codes]
From \eqref{pmat}, it can be easily observed that for the PR code $\mathrm{PR}(k,n,p(x))$, the codeword associated with the message $\mathbf{b}$, satisfies the linear recurrence which is characterized by $p(x)$. That is for any codeword $\mathbf{c}$, $p_0c_0\oplus p_1c_{1}\oplus\cdots\oplus p_kc_{k}=0$. In other words, each codeword of a PR code is a subsequence of length $n$ of a maximum-length sequence ($m$-sequence) with connection polynomial $x^kp(1/x)$. The encoding circuit of the PR code can then be represented by a LFSR \cite{Fredricsson1975}. It is important to note that LFSRs with connection polynomials $p(x)$ and $x^{k}p(1/x)$ are backward version of each other and hence have identical subsequence statistics \cite{Wainberg1970}. Therefore, their equivalent PR codes have the same Hamming weight distributions. 
\end{remark}

\begin{remark}
\label{dualprlfsr}
The PR code with any primitive polynomial $p(x)$ of degree $k$ and truncated at length $n=2^k-1$ is equivalent to the dual of the binary Hamming code of codeword length $2^{k}-1$ and message length $2^k-k-1$ with generator polynomial $p(x)$, also referred to as the Simplex code. Further, a PR code $\mathrm{PR}(k,n,p(x))$ is a dual of a shortened Hamming code $(n,n-k)$, where all codewords corresponding to polynomials of degree greater than or equal to $n$ are deleted from the original Hamming code \cite{Fredricsson1975}. A PR code $\mathrm{PR}(k,n,p(x))$ can be also realized as a punctured Simplex code\footnote{In general, every linear code over $\mathbf{GF}(q)$ with dual distance at least 3 is a punctured code of a Simplex code over $\mathbf{GF}(q)$ \cite[Corollary 10]{liu2020shortened}. Moreover, every linear code with minimum distance at least 3 is a shortened code of a Hamming code over $\mathbf{GF}(q)$ \cite[Theorem 11]{liu2020shortened}.}. 
\end{remark}

\begin{remark}[Boolean function construction of PR codes]
Let $f:\mathbf{GF}(2^k)\to\mathbf{GF}(2^k)$ denote a linear Boolean function \cite{Ding2016}; that is for any $\beta$ and $\gamma$ from $\mathbf{GF}(2^k)$ and $v$ and $u$ from $\mathbf{GF}(2)$, we have $f(v\beta+u\gamma)=vf(\beta)+uf(\gamma)$ and $f(\mathbf{0})=0$. We develop a code using a primitive polynomial $p(x)$ and primitive element $\alpha$, that is for any information block $\mathbf{b}=\alpha^j$, the $i^{th}$ coded symbol is $c_i=f(\alpha^{j+i-1})$. The code is then equivalent to the PR code $\mathrm{PR}(k,p(x))$\footnote{A more generic construction of a PR code over $\mathbf{GF}(p)$ where $p$ is prime, can be explained as follows. Let $D=\{d_1,d_2,\cdots,d_n\}\subseteq\mathbf{GF}(p^k)$. A PR code of length $n$ over $\mathbf{GF}(p)$ is constructed by
    $\mathcal{C}_D=\{\left(\mathrm{Tr}(xd_1),\mathrm{Tr}(xd_2),\cdots,\mathrm{Tr}(xd_n)\right):~x\in\mathbf{GF}(p^k)\}$
where $D$ is called the defining set of the code and $\mathrm{Tr}$ denotes the trace function from $\mathbf{GF}(p^k)$ onto $\mathbf{GF}(p)$ \cite{Ding2016}.}. 
\end{remark}
\begin{IEEEproof}
It is easy to show that for this code we have $\sum_{i=0}^{k}p_if(\alpha^{j+i})=f(\sum_{i=0}^{k}p_i\alpha^{j+i})=0$, which follows directly from the fact that $f$ is linear and $p(\alpha)=0$. It is then straightforward that the parity check matrix of this code is \eqref{pmat}. This completes the proof.
\end{IEEEproof}

\begin{figure}[t]
  \centering
  \includegraphics[width=0.55\columnwidth]{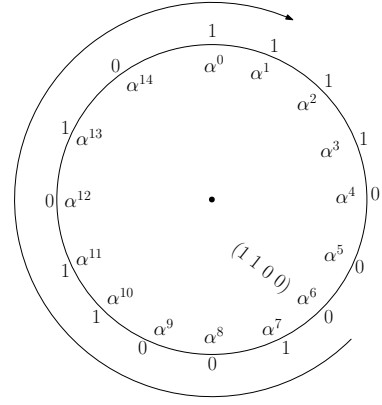}
  \caption{The encoding ring for a PR code generated by $p(x)=1+x+x^4$ and $f(\mathbf{b})=b_0\oplus b_1\oplus b_2 \oplus b_3$ and terminated at length $n=11$.}
  \label{figring}
\end{figure}

\begin{example} We assume that $k=4$ and the primitive polynomial is $p(x)=1+x+x^4$. The binary function $f$ is defined as $f(\mathbf{b})=b_0\oplus b_1\oplus b_2 \oplus b_3$. We list all non-zero elements of $\mathbf{GF}(2^4)$ over a circle as shown in Fig. \ref{figring} and calculate the binary value of each element subject to function $f$. The codeword associated with each message can be easily found by all values on a semi-ring started from the corresponding element of the field to the message vector and terminated at the desired length. For example, for the message $b=(1100)$ the codeword of length $n=11$ is $c=(01001101011)$. 
\end{example}

\begin{remark}
\label{remarkpoly}
Any PR code $\mathrm{PR}(k,n,p(x))$ is the dual of a polynomial code of codeword length $n$, message length $n-k$, and generator polynomial $p(x)$.
\end{remark}
\begin{IEEEproof}
The codewords of a polynomial code with generator polynomial $p(x)$ are obtained by $c(x)=b(x)p(x)$, where $b(x)=\sum_{i=0}^{n-k-1}b_ix^i$ is the message vector. The polynomial multiplication $b(x)p(x)$ is equivalent to the row-wise operation of $\mathbf{H}_n$. That is $c(x)=\sum_{i=0}^{n-k-1}b_ix^ip(x)$, which is equivalent to the product $\mathbf{b}\mathbf{H}_n$. In other words, $\mathbf{H}_n$ is the generator matrix of the polynomial code of codeword length $n$, message length $n-k$, and generator polynomial $p(x)$.
\end{IEEEproof}

\begin{lemma}
\label{lemma:disjoint}
Let $\mathrm{PR}(k,n,p_1(x))$ and $\mathrm{PR}(k,n,p_2(x))$ denote two PR codes that are generated with primitive polynomials $p_1(x)$ and $p_2(x)$, respectively, where $p_1(x)\ne p_2(x)$ and $n\ge 2k$. For simplicity of notations, we denote their respective codebooks by $\mathcal{C}_1$ and $\mathcal{C}_2$. Then, these codes do not have any non-zero codeword in common, i.e., $\mathcal{C}_1\bigcap \mathcal{C}_2=\{\mathbf{0}\}$.
\end{lemma}
\begin{IEEEproof}
Let us assume that $n=2k$ and $\mathbf{c}\ne \mathbf{0}$ is a codeword, which belongs to both $\mathcal{C}_1$ and $\mathcal{C}_2$. Therefore, $\mathbf{c}$ will satisfy parity check constraints for both $\mathcal{C}_1$ and $\mathcal{C}_2$. In particular, we have 
\begin{align}
    \left[\begin{array}{ccccccc}
    p_{1,0}&\cdots&p_{1,k-1}&p_{1,k}&0&\cdots&0\\
    &&&\vdots&&&\\
    0&\cdots&0&p_{1,0}&\cdots&p_{1,k-1}&p_{1,k}\\
    p_{2,0}&\cdots&p_{2,k-1}&p_{2,k}&0&\cdots&0\\
    &&&\vdots&&&\\
    0&\cdots&0&p_{2,0}&\cdots&p_{2,k-1}&p_{2,k}\\
    \end{array}\right]\mathbf{c}=\mathbf{0}_{2k}.
    \label{eq:bigH}
\end{align}
A linear combination of some rows of this matrix can be written as
\begin{align}
    g_1(x)p_1(x)+g_2(x)p_2(x),
\end{align}
where $g_1(x)$ and $g_2(x)$ are binary polynomials of degrees at most $k-1$. Let us assume that there are $g_1(x)$ and $g_2(x)$ such that $h(x)=g_1(x)p_1(x)+g_2(x)p_2(x)=0$. Therefore, $h(\alpha_1)=0$, where $\alpha_1$ is the primitive root of $p_1(x)$. We therefore have $g_2(\alpha_1)p_2(\alpha_1)=0$. Since $p_2(x)$ is a primitive polynomial other than $p_1(x)$, we have $p_2(\alpha_1)\ne0$, which results in $g_2(\alpha_1)=0$. This contradicts the fact that $p_1(x)$ is a primitive polynomial with primitive element $\alpha_1$, as the degree of $g_2(x)$ is less than $k$. Thus, every linear combination of rows of the parity check matrix above is non-zero. Therefore, the above parity check matrix is full-rank and the only solution to (\ref{eq:bigH}) is $\mathbf{c}=\mathbf{0}$. Since the first subsequence of length $2k$ of any non-zero codeword of $\mathcal{C}_1$ is  different than the first subsequence of length $2k$ of any non-zero codeword of $\mathcal{C}_2$, we can conclude that $\mathcal{C}_1\bigcap\mathcal{C}_2=\{\mathbf{0}\}$, for $n\ge 2k$.
\end{IEEEproof}

\begin{remark}
Lemma \ref{lemma:disjoint} can be generalized as follows. Any two PR codes $\mathrm{PR}(k_1,n,p_1(x))$ and $\mathrm{PR}(k_2,n,p_2(x))$ do not have any non-zero codeword in common, when $n\ge2\max\{k_1,k_2\}$ and $p_1(x)\ne p_2(x)$. This follows from the fact that the minimal polynomial\footnote{The minimal polynomial of sequence $\mathbf{c}$ is the connection polynomial of the shortest LFSR capable of producing $\mathbf{c}$ \cite{massey1969shift}.} of any codeword of length $n\ge 2k_1$ of $\mathrm{PR}(k_1,n,p_1(x))$ is unique and equals to $p_1(x)$ \cite{massey1969shift}. Similarly the minimal polynomial of any codeword of length $n\ge 2k_2$ of $\mathrm{PR}(k_2,n,p_2(x))$ is unique and equals to $p_2(x)$. However, if a non-zero codeword $\mathbf{c}$ of length $n\ge\max\{k_1,k_2\}$ belongs to both codebooks, it will have two distinct minimal polynomials, which contradicts with the uniqueness of the minimal polynomial.
\end{remark}
\begin{lemma}
\label{lemma:euler}
There are $\frac{\varphi(2^k-1)}{k}$ PR codes of message length $k$, where $\varphi(.)$ is the Euler's totient function. 
\end{lemma}
\begin{IEEEproof}
The number of primitive polynomials over $\mathbf{GF}(2)$ having degree $k$ is given by $\varphi(2^k-1)/k$ \cite{Helleseth1991}. Each primitive polynomial generates a PR code. This completes the proof.
\end{IEEEproof}

\section{Hamming Weight Distribution of Primitive Rateless Codes}
In this section, we will characterize the Hamming weight distribution of PR codes. Authors in \cite{Fredricsson1975} introduced an ideal distribution for the Hamming weight of the non-zero $n$-tuples of an $m$-sequence (equivalent to a PR code) as follows:
\begin{align}
    \hat{A}_t=\frac{2^k-1}{2^n-1}\dbinom{n}{t}, ~~~ 1\le t\le n,
    \label{eq:idealdist}
\end{align}
where they tried to characterize the deviation of the Hamming weight distribution of the $n$-tuples from the ideal distribution. In particular, they found that if the minimum Hamming weight of the shortened Hamming code is greater than $L$, the deviation from the ideal distribution is lower bounded by
\begin{align}
    ||A||^2\ge \frac{(1-2^{k-n})(2^{n-k}-1)}{(1-2^{-k})(2^n-1)}\sum_{i=1}^{L}\dbinom{n}{i},
\end{align}
where $||A||$ is the Euclidean norm of the gap between the Hamming weight distribution and the ideal distribution. The Hamming weight distribution of the $n$-tuples was also characterized, which relies on the Hamming weight distribution of the shortened Hamming code and thus cannot be scaled to moderate and large codes.  Moreover, according to Remark \ref{dualprlfsr}, a PR code $\mathrm{PR}(n,k,p(x))$ can be realized as a punctured Simplex code, for which the weight distribution at block lengths $n>2^{k-1}$ has been studied in \cite{baldi2012class}. The approach however cannot be extended for an arbitrary $n$  and in particular for $2k\le n\le 2^k-1$, which is the primary focus of this work.

Authors in \cite{Lidl1997} provided a bound for the weight of subsequences of a $m$-sequence.  That is for every $n$, we have \cite[Theorem 8.85]{Lidl1997}:
\begin{align}
    \left| w_{\mathrm{H}}(\mathbf{c})-\frac{n}{2}\right|\le\sqrt{k}\left(\frac{\log\left(2^k-1\right)}{\pi}+1\right),
\end{align}
and when $n$ goes large, we have ${w_\mathrm{H}(\mathbf{c})}\approx\frac{n}{2}$ \cite{wang2017quickest}, where ${w_\mathrm{H}(\mathbf{c})}$ is the Hamming weight of the subsequnce $\mathbf{c}$ of length $n$. This means that for a sufficiently large $n$, the Hamming weight distribution of PR codes is concentrated around $n/2$. The following lemma characterises the first and second moments of the Hamming weight distribution of PR codes. 

\begin{lemma}
\label{lemma:avgweigth}
A PR code $\mathrm{PR}(n,k,p(x))$ with $n\le2^k-1$ has the average Hamming weight equals to $\mu_n=\frac{n}{2}$ and the variance of the Hamming weights is $\sigma^2_n=\frac{n}{4}$\footnote{Lemma \ref{lemma:avgweigth} was previously presented in \cite{Lindholm1968}, which stated that the first and second moments of the distribution of the number of 1s in a subsequence of length $n$ of an $m$-sequence with primitive connection polynomial $p(x)$ of degree $k$, are $n/2$ and $n/4$, respectively. We however provide our proof for the completeness of the discussion.}.
\end{lemma}
\begin{IEEEproof}
The dual of the PR code $\mathrm{PR}(n,k,p(x))$ with $n\le2^k-1$ has a minimum Hamming weight of at least 3. This can be easily proved as $p(x)$ is primitive and does not have any binomial multiple with degree less than $2^k-1$. Because otherwise there would exist $0\le i_1<i_2<2^k-1$ such that $\alpha^{i_1}+\alpha^{i_2}=0$ for $\alpha$ being the root of $p(x)$. This implies that $\alpha^{i_2-i_1}=1$, which contradicts the fact that $\alpha$ is primitive. Let $A(z)=\sum_{i=0}^nA_iz^i$ and $B(z)=\sum_{i=0}^nB_iz^i$ denote the weight enumerator function of the PR code and its dual code, respectively. By using the MacWilliams identity \cite{MacWilliams1962}, we will have
\begin{align}
    B_1=\frac{1}{2^k}\sum_{j=0}^{n}A_j(n-2j),
\end{align}
which results in $\mu_n=\frac{1}{2^k}\sum_{j=0}^njA_j=\frac{n}{2}$, since $B_1=0$. Similarly, for $B_2=0$ we have
\begin{align}
    B_2=\frac{1}{2^k}\sum_{j=0}^{n}A_j\left(2\left(j-\frac{n}{2}\right)^2-\frac{n}{2}\right),
\end{align}
which results in $\sigma^2_n=\frac{1}{2^k}\sum_{j=0}^nA_j(j-\mu_n)^2=\frac{n}{4}$. 
\end{IEEEproof}

\begin{corollary}
The average Hamming weight of any PR code $\mathrm{PR}(k,n,p(x))$ increases by 1 when $n$ is increased by 2.
\end{corollary}

Let $A^{(i)}(z)$ denote the weight enumerator of the PR code $\mathrm{PR}(n,k,p_i(x))$. As we have $M_k=\varphi(2^k-1)/k$ of such codes (Lemma \ref{lemma:euler}), we can define the average weight enumerator of PR codes of dimension $k$ truncated at length $n$ as follows:
\begin{align}
    \Bar{A}(z)=\frac{1}{M_k}\sum_{i=1}^{M_k}\sum_{j=0}^{n}A^{(i)}_jz^j,
\end{align}
where for simplicity, we define $\Bar{A}_j=\frac{1}{M_k}\sum_{i=1}^{M_k}A^{(i)}_j$. By using the MacWilliams identity \cite{MacWilliams1962}, we will have
\begin{align}
   \Bar{A}_j= \frac{1}{M_k}\sum_{i=1}^{M_k}\sum_{t=0}^{n}\frac{B^{(i)}_t}{2^{n-k}}K_j(t)
   =\frac{1}{2^{n-k}}\sum_{t=0}^{n}\Bar{B}_tK_j(t),
   \label{eq:mcwilliam}
\end{align}
where $\Bar{B}_t=\frac{1}{M_k}\sum_{i=1}^{M_k}B^{(i)}_t$ and
\begin{align}
    K_j(t)=\sum_{\ell=0}^{j}(-1)^{\ell}\dbinom{t}{\ell}\dbinom{n-t}{j-\ell}
\end{align}
is the Krawtchouk polynomial \cite{BenHaim2006}, where $t$ is an integer, $0\le t\le n$. We commonly use the following identities in the rest of the paper  \cite{Krasikov1995}:

$K_j(0)=\dbinom{n}{j}$, $K_n(t)=(-1)^t$, and $K_j(n)=(-1)^n\dbinom{n}{j}$.

We are interested in the \emph{average} Hamming weight distribution of all ensembles of PR codes. For this we consider all possible $\varphi(2^k-1)/k$ ensembles of PR codes and their dual codes to characterize the average Hamming weight distributions at any desired codeword length. 
\subsection{Average Weight Distribution of Dual of PR Codes}
As stated in Remark \ref{remarkpoly}, the dual of a PR code is a polynomial code with generator polynomial $p(x)$. In other words, every codeword of the dual code is a product of $p(x)$. It is then clear that $\Bar{B}_t$ is equivalent to the expected number of all weight-$t$ multiples with degree at most $n-1$ of every primitive polynomial of degree $k$. The following lemma characterizes $\Bar{B}_t$.

\begin{lemma}
For the dual of PR codes of dimension $k$ and truncated at length $n$, the expected number of codewords of weight $t$, for $3\le t\le n$, is given by: 
\begin{align}
    \bar{B}_t\approx\frac{N_{k,t}}{\dbinom{2^k-2}{t-1}}D^{(k)}_{n,t},
    \label{eq:dualweigth}
\end{align}
where 
\begin{align}
D^{(k)}_{n,t}=\sum_{c=\max(k,t-1)}^{n-1}\dbinom{c-1}{t-2}(n-c),
\end{align}
and
\begin{align}
    N_{k,t}=\frac{\dbinom{2^k-2}{t-2}-N_{k,t-1}}{t-1}-\frac{\left(2^k-t+1\right)N_{k,t-2}}{t-2},
    \label{eq:tnomial}
\end{align}
and $N_{k,2}=N_{k,1}=0$.
\end{lemma}
\begin{IEEEproof}
Let $N_{k,t}$ denotes the number of $t$-nomial multiples (having constant term 1) of a primitive polynomial $p(x)$ of degree $k$, with initial condition $N_{k,2}=N_{k,1}=0$. It was shown in \cite{Maitra2002} that $N_{k,t}$ can be precisely characterized by (\ref{eq:tnomial}). It was further elaborated in \cite{Gupta2001} that the distribution of $t$-nomial multiples of degree less than or equal to $2^k-2$ is very close to the distribution of all distinct ($t-1$) tuples from $1$ to $2^k-2$. Under this assumption, referred to as \emph{Random Estimate} in \cite{Gupta2001}, the probability that a randomly chosen $t$-nomial of degree at most $2^k-2$ is a multiple of a primitive polynomial is given by $N_{k,t}/{\dbinom{2^k-2}{t-1}}$ \cite{Gupta2001,Maitra2002,Venkateswarlu2002}. The expected number of $t$-nomial multiples having degree equals to $c$, for $c\ge\max(k,t-1)$ is then given by $N_{k,t}\dbinom{c-1}{t-2}/\dbinom{2^k-2}{t-1}$. This follows from the fact that there are exactly $\dbinom{c-1}{t-2}$ many $t$-nomials of degree $c$. 
It is also clear that when a $t$-nomial $r(x)=1+x^{i_1}+\cdots+x^{i_{t-2}}+x^{c}$ is a multiple of $p(x)$, then $x^{i}r(x)$ for $0\le i\le n-c-1$ is also a multiple of $p(x)$ and has weight $t$. There are $n-c$ of such multiples, where $\max\{k,t-1\}\le c\le n-1$. Therefore, the expected number of weight $t$ polynomials of maximum degree $n-1$, which are multiples of primitive polynomial $p(x)$, is given by (\ref{eq:dualweigth}).
\end{IEEEproof}
Authors in \cite{Gupta2001} further approximated $N_{k,t}$ by $\dbinom{2^k-2}{t-2}/(t-1)$, which is tight when $t\ll 2^k$. By using this approximation, we can further simplify (\ref{eq:dualweigth}) as follows:
\begin{align}
    \bar{B}_t\approx\frac{1}{2^k-t}D^{(k)}_{n,t},~~3\le t\le n.
    \label{eq:dualweigth2}
\end{align}
\begin{prop}
\label{prop:1}
For $t\ge k+1$, we have $D^{(k)}_{n,t}=\dbinom{n}{t}$.
\end{prop}
\begin{IEEEproof}
Let $\xi:=c-1$ and $\eta:=t-2$, we then have
\begin{align}
    \nonumber D^{(k)}_{n,t}&=\sum_{c=t-1}^{n-1}\dbinom{c-1}{t-2}(n-c)=\sum_{\xi=\eta}^{n-2}\dbinom{\xi}{\eta}\dbinom{n-1-\xi}{t-1-\eta}\\
    \nonumber &\overset{(a)}{=}\sum_{\xi=0}^{n-1}\dbinom{\xi}{\eta}\dbinom{n-1-\xi}{t-1-\eta}-\dbinom{n-1}{\eta}\dbinom{0}{t-1-\eta}\\
    \nonumber&~~~-\sum_{\xi=0}^{\eta-1}\dbinom{\xi}{\eta}\dbinom{n-1-\xi}{t-1-\eta}\\
    &\overset{(b)}{=}\dbinom{n}{t},
\end{align}
where step $(a)$ expands the summation into three overlapping terms, and step $(b)$ follows from the fact that $\dbinom{0}{t-1-\eta}=0$, and for  $\eta>\xi$, we have
\begin{align}
    \dbinom{\xi}{\eta}=\frac{\xi\times(\xi-1)\cdots\times0\times\cdots(\xi-\eta+1)}{\eta!}=0,
\end{align}
and $\sum_{\xi=0}^{n-1}\dbinom{\xi}{\eta}\dbinom{n-1-\xi}{t-1-\eta}=\dbinom{n-1+1}{t-1+1}$ due to the Chu–Vandermonde identity \cite{Askey1975}.
\end{IEEEproof}
\begin{prop}
\label{prop:2}
For $t\le k$, $D^{(k)}_{n,t}$ is given by
\begin{align}
    D^{(k)}_{n,t}=\dbinom{n}{t}-\sum_{c=t-1}^{k-1}\dbinom{c-1}{t-2}(n-c).
    \label{eq:dktb}
\end{align}
\end{prop}
\begin{IEEEproof}
For $t\le k$, we have
\begin{align}
    \nonumber D&^{(k)}_{n,t}=\sum_{c=k}^{n-1}\dbinom{c-1}{t-2}(n-c)\\
    &=\sum_{c=t-1}^{n-1}\dbinom{c-1}{t-2}(n-c)-\sum_{c=t-1}^{k-1}\dbinom{c-1}{t-2}(n-c),
\end{align}
which reduces to (\ref{eq:dktb}) following the proof of Proposition \ref{prop:1}.
\end{IEEEproof}

\begin{figure}[t]
    \centering
    \includegraphics[width=\columnwidth]{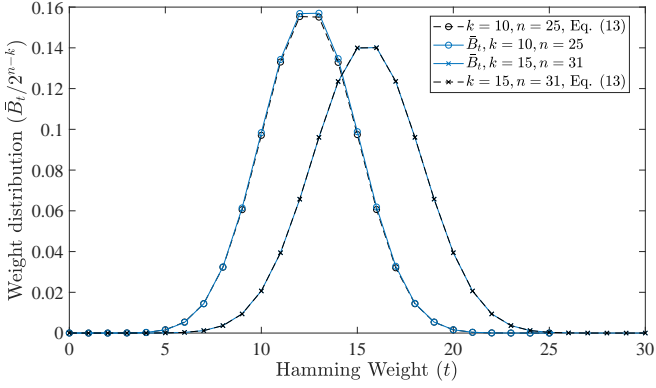}
    \vspace{-2ex}
    \caption{The average weight enumerator $\Bar{B}_t$ of duals of PR codes. Solid and dashed curves show the exact and approximate weight distributions, respectively.}
    \label{fig:avgweightdual}
\end{figure}

Fig. \ref{fig:avgweightdual} shows the average Hamming weight distribution of the dual of PR codes when $k=10$, $n=25$ and $k=15$, $n=31$. It is important to note that there are $M_{10}=60$ and $M_{15}=1800$ different PR codes of dimension $k=10$ and $k=15$, respectively. Fig. \ref{fig:avgweightdual} is produced by generating all the dual codes and their weight distributions and taking average to find the average weight distribution. As can be seen in this figure, (\ref{eq:dualweigth}) provides a tight approximation for the average weight distribution of duals of PR codes.

\begin{figure}[t]
    \centering
    \includegraphics[width=\columnwidth]{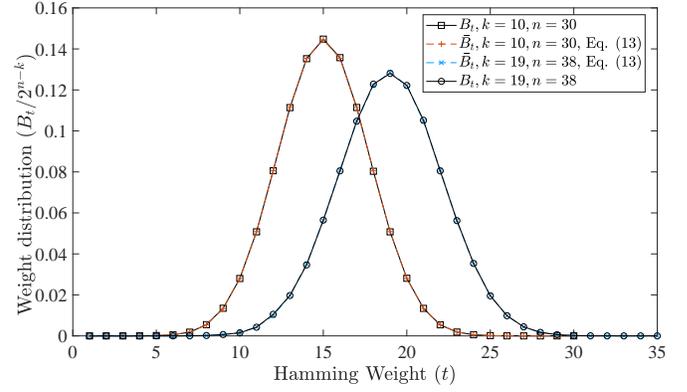}
    \vspace{-2ex}
    \caption{The weight distribution of the dual of the PR code. The primitive polynomials for $k=10$ and $k=19$ are respectively $p(x)=1+x^2+x^4+x^5+x^7+x^9+x^{10}$ and $p(x)=1+x+x^2+x^5+x^7+x^8+x^{12}+x^{13}+x^{14}+x^{17}+x^{19}$.   }
    \label{fig:weightdual}
\end{figure}

Fig. \ref{fig:weightdual} shows the Hamming weight distribution of the dual of a PR code at different lengths, when $k=10$ and $k=19$, with primitive polynomials $p(x)=1+x^2+x^4+x^5+x^7+x^9+x^{10}$ and $p(x)=1+x+x^2+x^5+x^7+x^8+x^{12}+x^{13}+x^{14}+x^{17}+x^{19}$, respectively. As can be seen,  (\ref{eq:dualweigth}) also provides a very tight approximation of the Hamming weight distribution of duals of PR codes. To better characterize the approximation in (\ref{eq:dualweigth2}), we use the Kullback-Leibler Divergence (KLD)\footnote{For discrete probability distributions $P$ and $Q$ defined on the same probability space, $\mathcal{X}$, the KLD (or relative entropy) from $Q$ to $P$ is defined to be $\mathrm{KLD}(P \parallel Q) = \sum_{x\in\mathcal{X}} p(x) \log\left(p(x)/q(x)\right)$ \cite{MacKay2003}.} to measure the distance between the exact weight distribution of the dual code and the approximations. In particular, when $k=10$, we have $\mathrm{KLD}(B_t\parallel\Bar{B}_t)=1.1\times 10^{-5}$ and when $k=19$, we have $\mathrm{KLD}(B_t\parallel\Bar{B}_t)=6.7\times 10^{-5}$.

\subsection{Average Weight Distribution of PR Codes}
The following lemma characterises the average Hamming weight distribution of PR codes.
\begin{lemma}
\label{lem:WeightEnum}
The average number of codewords of Hamming weight $j\ge 3$ of all PR codes $\mathrm{PR}(k,n,p(x))$ is approximated by:
\begin{align}
\Bar{A}_j\approx-2^{-n}\sum_{t=0}^{k}F^{(k)}_{n,t}K_j(t),
\label{eq:avgWeightPR}
\end{align}
where $F^{(k)}_{n,t}$ is given below
\begin{align}F^{(k)}_{n,t}=\left\{\begin{array}{ll}
1-2^k;&t=0,\\
\dbinom{n}{t};&t=1,2,\\
\sum_{c=t-1}^{k-1}\dbinom{c-1}{t-2}(n-c);&3\le t\le k.\end{array}\right.\label{eq:fnkt}\end{align} 
\end{lemma}
\begin{IEEEproof}
By using the MacWilliams Identity (\ref{eq:mcwilliam}) \cite{MacWilliams1963}, the approximation for $\Bar{B}_t$ in (\ref{eq:dualweigth2}), and the fact that $\Bar{B}_0=1$ and $\Bar{B}_1=\Bar{B}_2=0$ (see the proof of Lemma \ref{lemma:avgweigth}), we have
\begin{align}
\nonumber\Bar{A}_j&\approx \frac{1}{2^{n-k}}K_j(0)+\frac{1}{2^{n-k}}\sum_{t=3}^{n}\frac{1}{2^k-t}D^{(k)}_{n,t}K_j(t)\\
\nonumber&\overset{(a)}{\approx}\frac{1}{2^{n-k}}K_j(0)+\frac{1}{2^{n-k}}\sum_{t=3}^{n}\frac{1}{2^k}D^{(k)}_{n,t}K_j(t)\\
\nonumber&\overset{(b)}{=}\frac{1}{2^{n-k}}K_j(0)+\frac{1}{2^n}\sum_{t=0}^{n}\dbinom{n}{t}K_j(t)-\frac{1}{2^n}\sum_{t=0}^{2}\dbinom{n}{t}K_j(t)\\
\nonumber&-\frac{1}{2^n}\sum_{t=3}^{k}K_j(t)\sum_{c=t-1}^{k-1}\dbinom{c-1}{t-2}(n-c)\\
\nonumber&\overset{(c)}{=}\delta_j-\frac{1-2^k}{2^n}K_j(0)-\frac{1}{2^n}\sum_{t=1}^{2}\dbinom{n}{t}K_j(t)\\
\nonumber&-\frac{1}{2^n}\sum_{t=3}^{k}K_j(t)\sum_{c=t-1}^{k-1}\dbinom{c-1}{t-2}(n-c)\\
&\overset{(d)}{=}\delta_j-2^{-n}\sum_{t=0}^{k}F^{(k)}_{n,t}K_j(t).
\end{align}
where step $(a)$ follows from the fact that $t\le n\ll 2^k$, step $(b)$ follows from Proposition \ref{prop:1} and Proposition \ref{prop:2}, step $(c)$ follows from $\sum_{t=0}^{n}\dbinom{n}{t}K_j(t)=2^n\delta_j$ \cite{BenHaim2006}, where $\delta_j$ is the Kronecker delta function, i.e., $\delta_0=1$ and $\delta_j=0$ for $j>0$, and step $(d)$ follows from the definition of $F^{(k)}_{n,t}$ in \eqref{eq:fnkt}.
\end{IEEEproof}
Following Lemma \ref{lem:WeightEnum}, the average weight enumerator of PR codes of dimension $k$ and truncated at length $n$ is given by:
\begin{align}
    \Bar{A}(z)\approx 1-{2^{-n}}\sum_{t=0}^{k}F^{(k)}_{n,t}(1-z)^t(1+z)^{n-t}.
\end{align}

\begin{figure}
    \centering
    \includegraphics[width=\columnwidth]{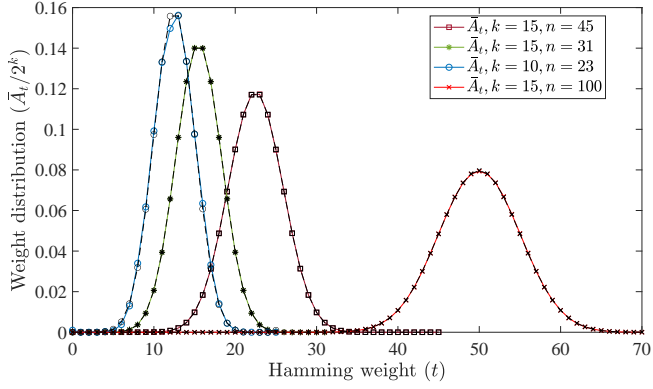}
    \vspace{-2ex}
    \caption{The average weight distribution $\Bar{A}_t$ of PR codes. Solid and dashed curves show the exact and approximate weight distributions, respectively. The approximate weight distribution is obtained via (\ref{eq:avgWeightPR}).}
    \label{fig:avgWeightPR}
\end{figure}

Fig. \ref{fig:avgWeightPR} shows the average weight distribution of PR codes of dimension $k=10$ and $k=15$ at different lengths. As can be seen, (\ref{eq:avgWeightPR}) provides a tight approximation for the average weight distribution of PR codes. 

\begin{figure}[t]
  \centering
  \includegraphics[width=\columnwidth]{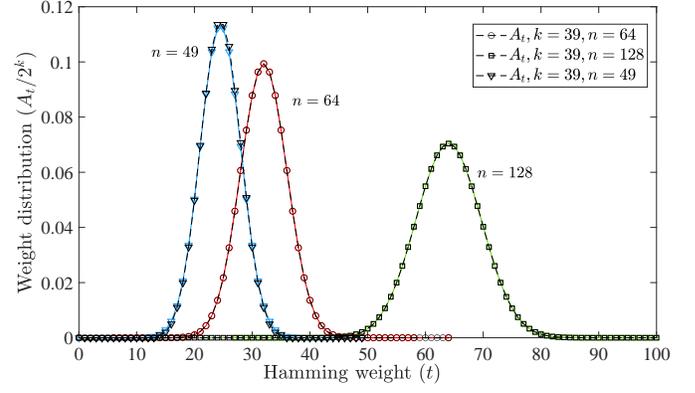}
  \vspace{-2ex}
  \caption{The weight distribution of a PR code with $k=39$ and $p(x)=x^{39} + x^{38} + x^{35} + x^{32} + x^{27} + x^{25} + x^{24} + x^{23} + x^{20} + x^{19} + x^{16} +    x^{15} + x^{11} + x^{9} + x^8 + x^4 + x^3 + x^2 + 1$. Solid and dashed lines respectively show the exact and approximate (obtained via (\ref{eq:avgWeightPR})) weight distributions.}
  \label{fig:weightdist39}
\end{figure}

Fig. \ref{fig:weightdist39} shows the Hamming weight distribution of a PR code when $k=39$ and $p(x)=x^{39} + x^{38} + x^{35} + x^{32} + x^{27} + x^{25} + x^{24} + x^{23} + x^{20} + x^{19} + x^{16} +    x^{15} + x^{11} + x^{9} + x^8 + x^4 + x^3 + x^2 + 1$. As can be seen, the wight distribution can be well approximated by (\ref{eq:avgWeightPR}). When $k=39$ and $n=49$, we have $\mathrm{KLD}(A_t\parallel\Bar{A}_{t})=8.6\times 10^{-11}$,  when $n=64$, $\mathrm{KLD}(A_t\parallel\Bar{A}_{t})=1.4\times 10^{-10}$, and when $n=128$, $\mathrm{KLD}(A_t\parallel\Bar{A}_{t})=5.4\times 10^{-10}$.  We used MAGMA calculator \cite{MR1484478} to obtain the Hamming weight distribution of PR codes. 

In the following theorem, we prove the existence of a PR code for any $k$ and $n\ge 2k$, that has a minimum Hamming weight large than a certain value. 
\begin{theorem}
For given $k$ and $n\ge2k$, there is at least one PR code with the minimum Hamming weight lower bounded by $d_{\min}$, where
\begin{align}
    d_{\min}=\max_d\left\{d\left|-2^{-n}\sum_{j=3}^{d}\sum_{t=0}^{k}F_{n,t}^{(k)}K_j(t)\le 1 \right.\right\}.
    \label{eq:mindisbound}
\end{align}
\end{theorem}
\begin{IEEEproof}
Let $d_{\min}=\max_d\{d|\sum_{j=3}^d\Bar{A}_j\le 1\}$.  Since $\Bar{A}_j=\frac{1}{M_k}\sum_{i=1}^{M_k} A_j^{(i)}$ and $M_k=\varphi(2^k-1)/k$, we have 
\begin{align}
    \sum_{j=3}^{d_{\min}}\sum_{i=1}^M A_j^{(i)}\le M_k.
\end{align}
This means that the total number of codewords with Hamming weight less than or equal to $d_{\min}$ of all PR codes of dimension $k$ and truncated at length $n$ is less than $M_k$. Since the sets of non-zero codewords of any two PR codes of dimension $k$ and truncated at length $n\ge 2k$ are disjoint (Lemma \ref{lemma:disjoint}), there should be at least one PR code of dimension $k$ and length $n$ that has a minimum Hamming weight larger than or equal to $d_{\min}$. This completes the proof.
\end{IEEEproof}

\begin{remark}
\label{rem:gv}
The average number of codewords of Hamming weight $j\ge 3$ of all PR codes of dimension $k$ truncated at length $n$ is upper bounded by:
\begin{align}
    \Bar{A}_j\le \frac{1}{2^{n-k}}\dbinom{n}{j}.
    \label{eq:binoapp}
\end{align}
\end{remark}
This can be verified from $\Bar{A}_j$ for $j\ge3$ in (\ref{eq:avgWeightPR}) as follows:
\begin{align}
\nonumber\bar{A}_j&=\frac{2^k-1}{2^{n}}K_j(0)-\frac{1}{2^{n}}\sum_{t=1}^k
F_{n,t}^{(k)}K_j(t)\\
&\overset{(a)}{=}\frac{1}{2^{n-k}}\dbinom{n}{j}-\frac{1}{2^{n}}\left(\dbinom{n}{j}+\sum_{t=1}^k
F_{n,t}^{(k)}K_j(t)\right),
\end{align}
where step $(a)$ follows from the fact that $K_j(0)=\dbinom{n}{j}$. When $n$ is sufficiently large, we have 
\begin{align}
    x_1^{(j)}=\frac{n}{2}-\sqrt{j(n-j)}+o(n),
\end{align}
where $x_1^{(j)}$ is the smallest root of $K_j(x)$. For $n$ sufficiently large and $j$ sufficiently small, we will have $K_j(t)\ge0$; therefore, $\Bar{A}_j\le 2^{k-n}\dbinom{n}{j}$.  It is important to note that as can be seen in Fig. \ref{fig:avgWeightPR} and Fig. \ref{fig:weightdist39}, the Hamming weight distribution of PR codes can be well approximated by the ideal distribution \eqref{eq:idealdist}, which is similar to the truncated binomial distribution \eqref{eq:binoapp}\footnote{Authors in \cite[Eq. 38]{Jordan1973} tried to compare the probability that there are exactly $t$ ones in $n$ successive bits of an $m$-sequence and the ideal distribution. The approach, however, depends on the primitive polynomial used to generate the $m$-sequence and is computationally complex when $k$ is large.}.

According to Remark \ref{rem:gv}, one can conclude that
\begin{align}
    d_{\min}\ge d_{\mathrm{GV}},
    \label{eq:boundgvmin}
\end{align}
where $d_{\mathrm{GV}}$ is the minimum Hamming weight obtained from the Gilbert-Varshamov bound \cite{Jiang2004}:
\begin{align}
    d_{\mathrm{GV}}=\max_d\left\{d\left|\sum_{j=0}^{d-1}\dbinom{n}{j}\le 2^{n-k} \right.\right\}.
    \label{eq:GVdisbound}
\end{align}

\begin{figure}
    \centering
    \includegraphics[width=\columnwidth]{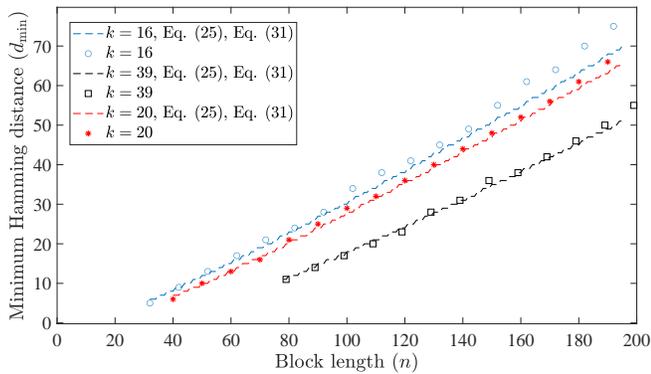}
    \vspace{-2ex}
    \caption{The minimum Hamming weight of PR codes of dimension $k=16$, $k=20$, and $k=39$ at different block lengths. The primitive polynomials are taken from Table \ref{tab:primpoly} and minimum Hamming weights were calculated by MAGMA \cite{MR1484478}.}
    \label{fig:mindistbound}
\end{figure}

\begin{figure*}
     \centering
       \subfloat[Low density primitive polynomial \label{fig:wdPlow}]{%
       \includegraphics[width=0.325\textwidth]{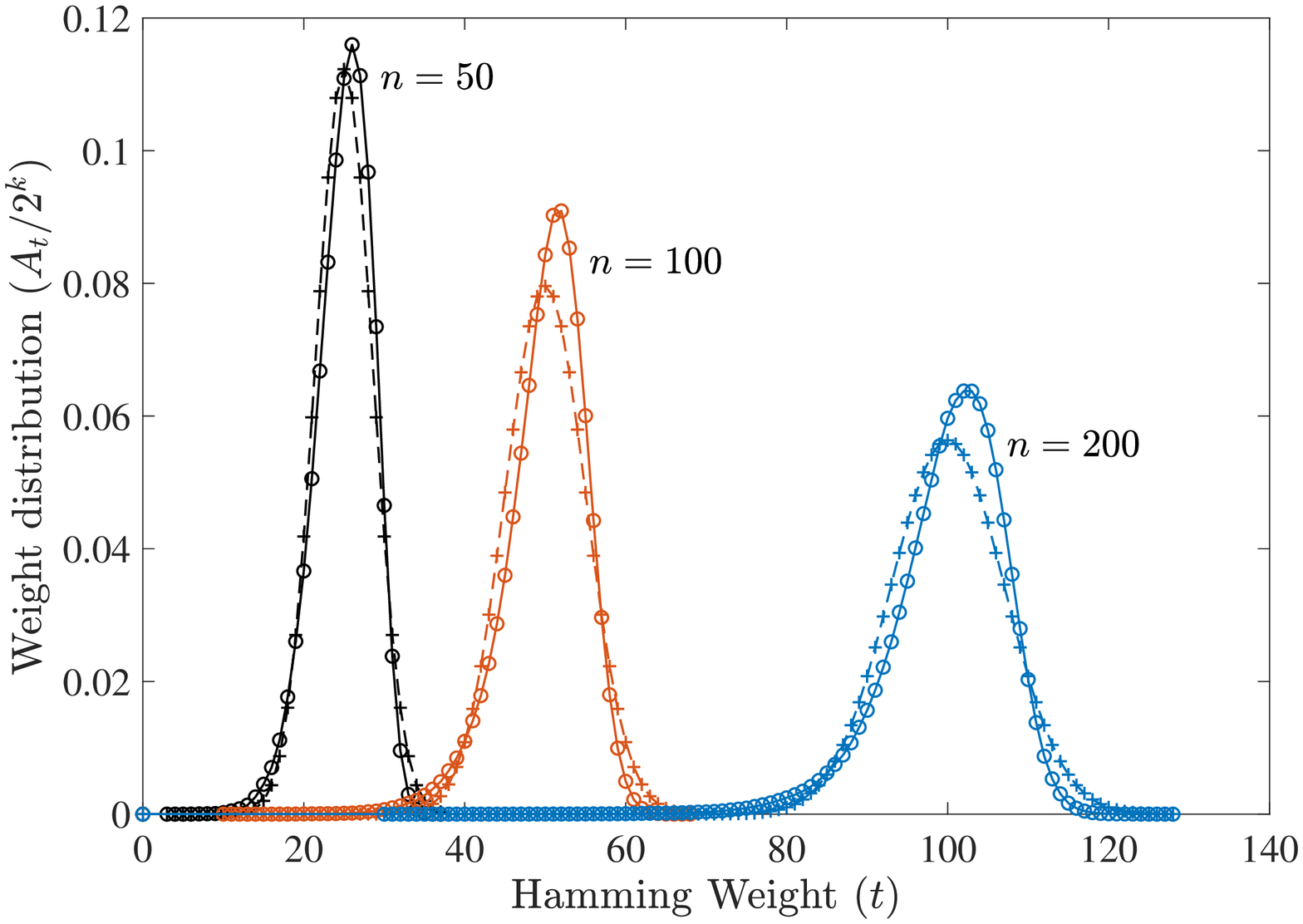}}
    \hfill
    \subfloat[Moderate density primitive polynomial \label{fig:wdPMod}]{%
       \includegraphics[width=0.325\textwidth]{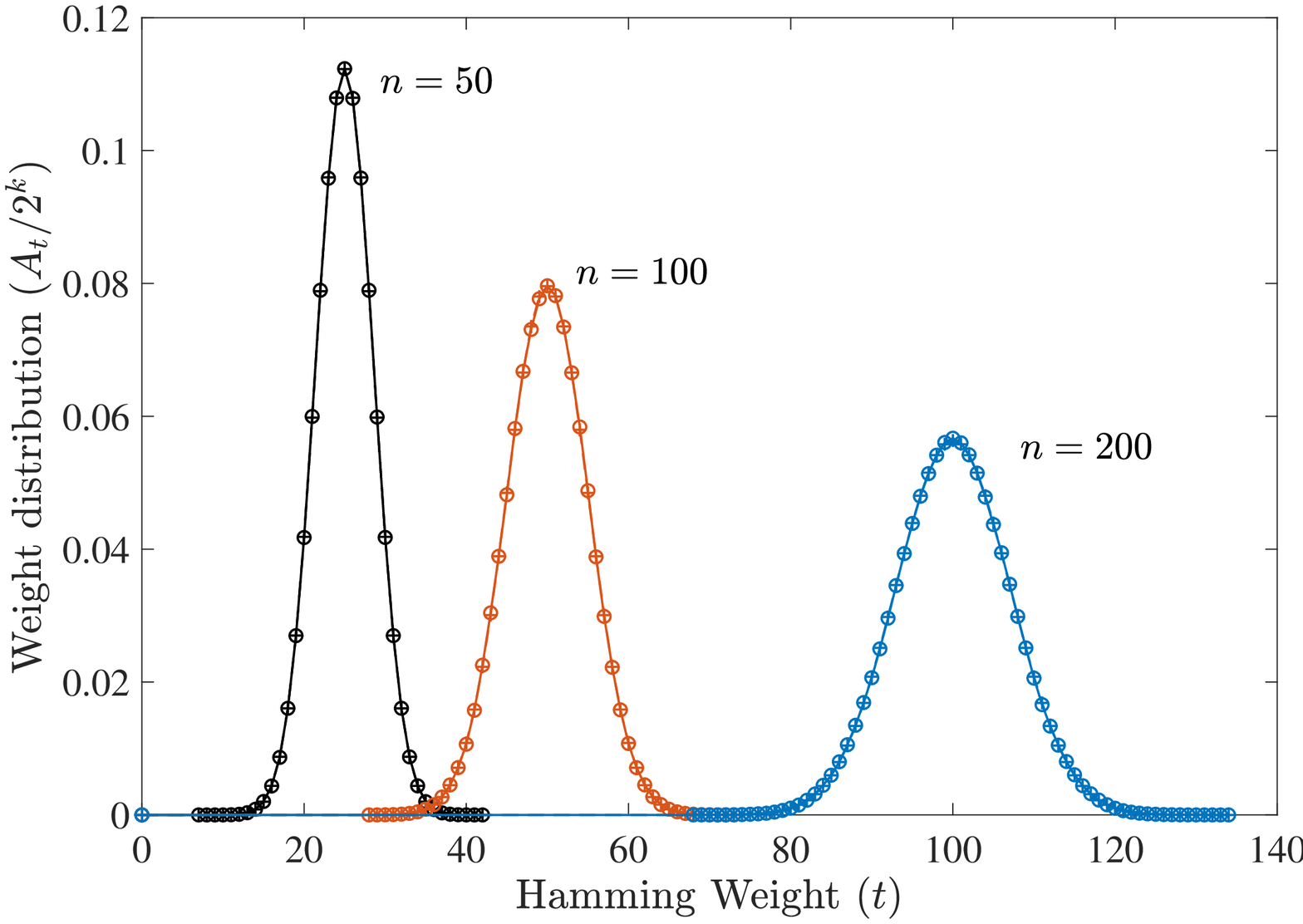}}
    \hfill
    \subfloat[High density primitive polynomial \label{fig:wdPHigh}]{%
       \includegraphics[width=0.325\textwidth]{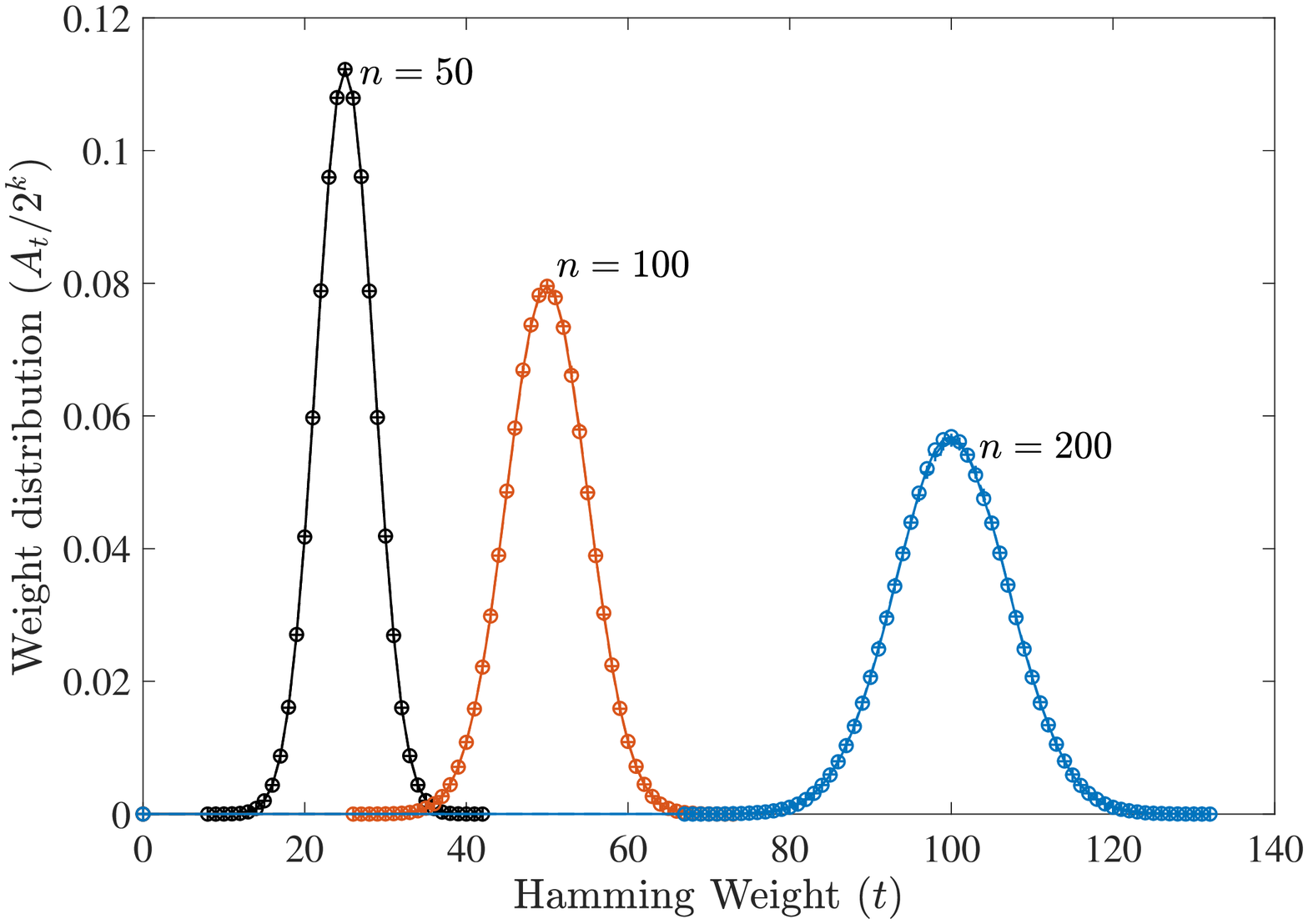}}
    \hfill
        \caption{The Hamming weight distribution of PR code when $k=23$ and a) $p(x)=1+x^5+x^{23}$, b) $p(x)=1+x^2+x^3+x^5+x^9+x^{10}+x^{12}+x^{14}+x^{16}+x^{18}+x^{23}$, and c) $p(x)=1+x^2+x^4+x^5+x^6+x^8+x^9+x^{10}+x^{11}+x^{15}+x^{16}+x^{17}+x^{18}+x^{19}+x^{21}+x^{22}+x^{23}$. Solid and dashed lines show the Hamming weight distribution and binomial distribution, respectively.}
        \label{fig:wdLMH}
\end{figure*}
Fig. \ref{fig:mindistbound} shows the minimum Hamming weight for three PR codes of dimension $k=16$, $k=20$, and $k=39$ at different block length. As can be seen in this figure, a PR code with a properly chosen primitive polynomial will have a minimum Hamming weight close to  the bound (\ref{eq:mindisbound}) at any block length $n\ge2k$. It is also important to note that the bound obtained in (\ref{eq:mindisbound}) and (\ref{eq:GVdisbound}) are identical for the cases in Fig. \ref{fig:mindistbound}. One can search for other primitive polynomials to achieve higher minimum Hamming weight at a given block length. This also shows that a PR code with a properly chosen primitive polynomial can meet the Gilbert-Varshamov bound.

In this paper, we use MAGMA \cite{MR1484478} to calculate the minimum Hamming weight of PR codes. MAGMA is using an algorithm described in \cite{betten2013codierungstheorie} to find the minimum Hamming weight of linear codes. It generates several generator matrices for the same code, such that these codes have disjoint information sets. The algorithm proceeds by enumerating all combinations derived from $r$ information symbols in all generator matrices, for each successive $r$. Once the lower and upper bounds on the minimum weight meet, the computation is complete. For PR codes, the codeword with the minimum weight usually corresponds to a low-weight message word; thus, MAGMA easily finds the minimum weight of PR codes.

\section{Selecting Primitive Polynomials for PR codes}
S. Wainberg and J. K Wolf \cite{Wainberg1970} studied the properties of subsequences of long $m$-sequences using the moments of the subsequence weight distribution. The moments were used for selecting good $m$-sequences for correlation-detection problem. Authors in \cite{Matsumoto1996} showed that for some bad initial vectors, terrible non-randomness continues for extraordinary long time in the sequences generated by the LFSRs with primitive polynomials with three terms. The bad behaviour of primitive trinomials and $t$-nomials with small $t$ was studies in \cite{Compagner1991,Fredricsson1975,Lindholm1968}. 

For PR codes to have the Hamming weight distribution closely approach the binomial distribution and accordingly a minimum Hamming weight lower bounded by \eqref{eq:boundgvmin}, the subsequences of the LFSR should preserve randomness for almost all initial conditions. Otherwise, the subsequences will have too many zeros or ones. Therefore, the primitive polynomial should be chosen properly to preserve randomness for subsequences of moderate length. Authors in \cite{Jordan1973} showed (via numerical results) that the probability distribution of the number of ones in $n$ successive bits of an $m$-sequence with primitive polynomial $p(x)$, which is equivalent to the Hamming weight distribution of the PR code with the same primitive polynomial, can be well approximated by the binomial distribution, when 1) the length $k$ of the shift-register is sufficiently large, 2) the sequence length $n$ satisfies $144\le n\ll 2^k$, and 3) the primitive polynomial should be chosen such that the number of subsequences of length $n$ with Hamming weight $\ell$ is small for $\ell$ near zero and $\ell$ near $n$ \cite{Jordan1973}. We however found that in most cases when the primitive polynomial is chosen properly, the Hamming weight distribution closely approaches the binomial distribution for a sufficiently large $n$. 

LFSRs with their connection polynomials very sparse are very vulnerable to various known attacks. On the other hands, a very dense primitive polynomial might be a factor of a low density polynomial of moderate degree, which makes the LFSR vulnerable to various attacks, such as the correlation attack \cite{Jambunathan2000}. When the primitive polynomial is sparse or has a multiple with only a few non-zero elements, the dual of the PR code will have a relatively low minimum Hamming weight. In particular, as shown in \cite{Matsumoto1996} when the primitive polynomial is $p(x)=1+x^{\ell}+x^k$, with $k\ge 2\ell$, the initial vector for the characteristic sequence has at most two 1's, where the characteristic sequences satisfies $x_i=x_{2i}$ for every integer $i$ \cite{Golomb1967}. This means that the sequence is not completely random; therefore, the Hamming weight distribution of the respective PR code deviates from the binomial distribution. 

\begin{table}[t]
    \centering
    \caption{The KLD of the Hamming weight distribution and the binomial distribution, $f_{\mathrm{bin}}(n,d)=\dbinom{n}{d}2^{-n}$.}
    \label{tab:kldLMH}
    \scriptsize
    \begin{tabular}{|c|c|c|c|c|c|c|}
    \hline
         \multirow{2}{*}{$n$} & \multicolumn{2}{c|}{Low-density $p(x)$} & \multicolumn{2}{c|}{Moderate-density $p(x)$}& \multicolumn{2}{c|}{High-density $p(x)$}\\
         \cline{2-7}        
         & $\mathrm{KLD}$&$d_{\min}$& $\mathrm{KLD}$&$d_{\min}$& $\mathrm{KLD}$&$d_{\min}$ \\
         \hline
         \hline
         50&$2.53e{-2}$&3& $7.29e{-6}$&7&$5.24e{-6}$&8\\
         \hline
         100&$5.07e{-2}$&10&$2.55e{-5}$&28&$2.46e{-5}$&26\\
         \hline
         200&$8.72e{-2}$&30&$3.84e{-5}$&68&$9.58e{-5}$&67\\
         \hline
    \end{tabular}
\end{table}
\begin{table*}
\centering
\caption{Some of good primitive polynomials for PR codes and their minimum Hamming weights at different rates. The block length at each rate is set to $n=\lceil k/R\rceil$, where $\lceil .\rceil$ is the ceiling operand. The MAGMA calculator \cite{MR1484478} was used to obtain the primitive polynomials and the minimum Hamming weights.}
\label{tab:primpoly}
\scriptsize
\begin{tabular}{|c|p{8.5cm}|c|c|c|c|c|c|}
\hline
    \multirow{2}{*}{$k$} &\multirow{2}{*}{$p(x)=1+x^k+$}& \multicolumn{6}{c|}{$d_{\min}$ ($d_{\mathrm{GV}}$)} \\
    \cline{3-8}
    &&$R=0.6$&$R=0.5$&$R=0.4$&$R=0.3$&$R=0.2$&$R=0.1$\\
    \hline
    \hline
    2&$x$&2 (3)&2 (3)&3 (3)&4 (4)&6 (5)&13  (10)\\
    \hline
    3&$x$&2 (3)&3 (3)&4 (4)&5 (5)&8 (7)&16  (14)\\
    \hline
    4&$x$&2 (3)&3 (3)&4 (4)&7 (6)&9 (9)&20  (19)\\
    \hline
    5&$x^2$&2 (3)&3 (3)&4 (5)&6 (6)&11 (10)&24  (23)\\
    \hline
    6&$x^2+x^3+x^5$&2 (3)&3 (4)&5 (5)&6 (7)&11 (11)&29  (26)\\
    \hline
    7&$x^1+x^3+x^6$&2 (3)&4 (4)&6 (5)&9 (7)&13 (12)&30  (29)\\
    \hline
    8&$x^2+x^3+x^5$&3 (4)&4 (4)&5 (5)&8 (8)&14 (13)&32  (29)\\
    \hline
    9&$x+x^3+x^4$&3 (4)&4 (4)&6 (6) &10 (8)&16 (14) &36 (32)\\
    \hline
    10&$x+x^2+x^3+x^5+x^6$&3 (4)&5 (4)&6 (6)&10 (9)&16 (15)&39 (36)\\
    \hline
    11&$x+x^3+x^5$&3 (4)&5 (5)&7 (7)&10 (10)&19 16)&43 (39)\\
    \hline
    12&$x+x^3+x^4+x^5+x^6$&3 (4)&5 (5)&7 (7)&10 (10)&17 (18)&45 (42)\\
    \hline
    13&$x+x^3+x^4+x^5+x^8$ &4 (4)&6 (5)&8 (7)&13 (11)&20 (19)&48 (45)\\
    \hline
    14&$x+x^3+x^5+x^6+x^7$&4 (4)&6 (5)&8 (8)&13 (12)&21 (20)&51 (48)\\
    \hline
    15&$x+x^2+x^5+x^7+x^8$&4 (4)&6 (6)&8 (8)&13 (12)&23 (21)&56 (52)\\
    \hline
    16&$x+x^4+x^6+x^8+x^9+x^{11}+x^{13}$&5 (4)&7 (6)&9 (8)&14 (13)&23 (23)&57 (55)\\
    \hline
    17&$x+x^2+x^3+x^6+x^{12}$&4 (5)&5 (6)&9 (9)&15 (14)&24 (24)&59 (58)\\
    \hline
    18&$x+x^2+x^4+x^6+x^7+x^8+x^9$&4 (5)&5 (6)&9 (9)&14 (14)&26 (25)&66 (61)\\
    \hline
    19&$x^2+x^3+x^4+x^5+x^6+x^7+x^8+x^9+x^{11}$&4 (5)&5 (6)&9 (10)&15 (15)&28 (26)&69 (64)\\
    \hline
    20&$x^2+x^3+x^4+x^7+x^{10}+x^{14}+x^{17}$&4 (5)&6 (7)&10 (10)&15 (16)&29 (28)&69 (68)\\
    \hline
    21&$x+x^2+x^3+x^6+x^7+x^{10}+x^{13}+x^{15}+x^{16}+x^{17}+x^{19}$&5 (5)&6 (7)&11 (10)&17 (16)&28 (29)&70 (71)\\
    \hline
    22&$x^7+x^{11}+x^{12}+x^{14}+x^{15}+x^{16}+x^{17}+x^{19}+x^{21}$&5 (5)&8 (7)&10 (11)&16 (17)&31 (30)&78(74)\\
    \hline
    23&$x^2+x^3+x^5+x^9+x^{10}+x^{12}+x^{14}+x^{16}+x^{18}$&5 (5)&6 (7)&12 (11)&18 (17)&32 (31)&78 (77)\\
    \hline
    24&$x+x^2+x^3+x^4+x^6+x^8+x^9+x^{14}+x^{21}+x^{22}+x^{23}$&5 (5)&7 (8)&11 (12)&17 (18)&33 (33)&82 (80)\\
    \hline
    25&$x^2+x^3+x^5+x^8+x^{11}+x^{15}+x^{16}+x^{17}+x^{18}+x^{20}+x^{21}+x^{23}+x^{24}$&5 (6)&7 (8)&12 (12)&21 (19)&35 (34)&84 (84)\\
    \hline
    26&$x^3+x^5+x^6+x^9+x^{13}+x^{14}+x^{16}+x^{17}+x^{19}+x^{24}+x^{25}$&5 (6)&7 (8)&12 (12)&20 (19)&35 (35)&91 (87)\\
    \hline
    27&$x^4+x^5+x^9+x^{12}+x^{15}+x^{16}+x^{18}+x^{22}+x^{24}+x^{25}+x^{26}$&6 (6)&9 (8)&13 (13)&21 (20)&37 (36)&94 (90)\\
    \hline
    28&$x+x^2+x^4+x^6+x^{10}+x^{11}+x^{16}+x^{19}+x^{21}+x^{22}+x^{23}+x^{25}+x^{26}$&5 (6)&9 (8)&13 (13)&20 (21)&38 (37)&93 (93)\\
    \hline
    29&$x^6+x^8+x^9+x^{10}+x^{12}+x^{18}+x^{22}+x^{24}+x^{26}$&6 (6)&8 (9)&14 (13)&21 (21)&40 (39)&100 (96)\\
    \hline
    30&$x^3+x^4+x^6+x^9+x^{11}+x^{12}+x^{16}+x^{18}+x^{23}+x^{26}+x^{27}$&5 (6)&9 (9)&14 (14)&22 (22)&42 (40)&103 (99)\\
    \hline
    31&$x+x^6+x^8+x^9+x^{12}+x^{13}+x^{18}+x^{19}+x^{21}+x^{22}+x^{25}+x^{26}+x^{27}+x^{28}+x^{30}$ &5 (6)&8 (9)&14 (14)&23 (23)&42 (41)&105 (100)\\
    \hline
    32&$x+x^2+x^5+x^7+x^8+x^9+x^{11}+x^{12}+x^{14}+x^{16}+x^{20}+x^{22}+x^{23}+x^{26}+x^{30}$&6 (7)&9 (9)&14 (14)&23 (23)&41 (42)&109 (106)\\
    \hline
    33&$x^2+x^3+x^4+x^5+x^7+x^9+x^{13}+x^{15}+x^{19}+x^{22}+x^{23}+x^{24}+x^{25}+x^{27}+x^{28}+x^{30}+x^{31}$&6 (7)&9 (10)&15 (15)&23 (24)&45 (44)&111 (109)\\
    \hline
    34&$x+x^5+x^8+x^{11}+x^{13}+x^{15}+x^{17}+x^{19}+x^{20}+x^{21}+x^{22}+x^{24}+x^{25}+x^{26}+x^{27}+x^{29}+x^{30}+x^{31}+x^{32}$ &6 (7)&10 (10)&15 (15)&25 (25)&45 (45)&116 (112)\\
    \hline
    35&$x^2+x^4+x^5+x^7+x^8+x^{10}+x^{11}+x^{12}+x^{13}+x^{14}+x^{16}+x^{18}+x^{19}+x^{21}+x^{23}+x^{25}+x^{27}+x^{29}+x^{34}$&7 (7)&9 (10)&15 (15)&26 (25)&47 (46)&118 (115)\\
    \hline
    36&$x^3+x^4+x^{10}+x^{12}+x^{13}+x^{14}+x^{15}+x^{16}+x^{17}+x^{21}+x^{22}+x^{23}+x^{24}+x^{26}+x^{27}+x^{28}+x^{29}+x^{30}+x^{31}+x^{32}+x^{34}$&6 (7)&9 (10)&16 (16)&26 (26)&48 (47)&120 (119)\\
    \hline
    37&$x^{5}+x^{9}+x^{12}+x^{13}+x^{16}+x^{17}+x^{19}+x^{20}+x^{21}+x^{22}+x^{25}+x^{26}+x^{27}+x^{28}+x^{31}+x^{32}+x^{33}$&7 (7)&10 (11)&16 (16)&26 (27)&49 (49)&124 (122)\\
    \hline
    38&$x+x^{4}+x^{5}+x^{7}+x^{11}+x^{14}+x^{16}+x^{18}+x^{19}+x^{21}+x^{23}+x^{26}+x^{27}+x^{28}+x^{29}+x^{31}+x^{35}$&6 (7)&10 (11)&16 (17)&28 (27)&52 (50)&122 (125)\\
    \hline
    39&$x+x^{2}+x^{3}+x^{4}+x^{6}+x^{8}+x^{10}+x^{12}+x^{13}+x^{14}+x^{16}+x^{18}+x^{20}+x^{26}+x^{29}+x^{32}+x^{33}+x^{34}+x^{38}$&7 (7)&11 (11)&17 (17)&28 (28)&52 (51)&128 (128)\\
    \hline
    40&$x+x^{4}+x^{6}+x^{7}+x^{8}+x^{10}+x^{12}+x^{15}+x^{16}+x^{17}+x^{19}+x^{20}+x^{21}+x^{24}+x^{25}+x^{26}+x^{27}+x^{30}+x^{31}+x^{32}+x^{33}+x^{35}+x^{37}+x^{38}+x^{39}$&8 (8)&11 (11)&18 (18)&29 (29)&52 (52)&131 (131)\\
    \hline
\end{tabular}
\end{table*}
\begin{example} Let $k=23$ and we consider three primitive polynomials, $p_1(x)=1+x^5+x^{23}$,  $p_2(x)=1+x^2+x^3+x^5+x^9+x^{10}+x^{12}+x^{14}+x^{16}+x^{18}+x^{23}$, and $p_3(x)=1+x^2+x^4+x^5+x^6+x^8+x^9+x^{10}+x^{11}+x^{15}+x^{16}+x^{17}+x^{18}+x^{19}+x^{21}+x^{22}x^{23}$. The first primitive polynomial is of low density and has only three terms. The second and third polynomials have respectively 11 and 17 terms. Fig. \ref{fig:wdPlow} shows the Hamming weight distribution of the PR code with $p_1(x)$ at block lengths $n=50, ~100$, and $200$. As can be seen, the weight distribution clearly deviates from the binomial distribution. However, as can be seen in Fig. \ref{fig:wdPMod} and Fig. \ref{fig:wdPHigh}, when the density of the primitive polynomial is moderate or high, the Hamming weight distribution of the PR code at different block lengths closely approach the binomial distribution. To better characterise the mismatch between the Hamming weight distribution and binomial distribution, we list the KLD between the distributions in Table \ref{tab:kldLMH}. As can be seen the KLD for PR codes with the moderate/high density primitive polynomial is significantly lower than that for PR codes with low-density primitive polynomials. It is also clear from Table \ref{tab:kldLMH} that the PR code with moderate to high density primitive polynomials achieve larger minimum Hamming weights.
\end{example}

We found some good primitive polynomials for PR codes which can closely achieve the bound \eqref{eq:mindisbound} developed in Theorem 1. In particular, first a random irreducible polynomial of degree $k$ and weight larger than or equal to $k/2$ is generated. Then, the polynomial will be tested for primitivity. If the polynomial is not primitive, another random irreducible polynomial will be generated. If the polynomial is primitive, then the minimum Hamming weight of the code is calculated at different rates. If the calculated Hamming weights are close to the bound in \eqref{eq:mindisbound}, then the polynomial will be considered as good primitive polynomial. Usually, the good polynomial is found after generating up to 5 random irreducible polynomials. When $k$ increases, the number of primitive polynomials of degree $k$ also increases, therefore, the search space for finding a better primitive polynomial also scales.

Table \ref{tab:primpoly} lists some of good primitive polynomials for PR codes of dimension up to $k=40$. For most of the primitive polynomials, the density of the primitive polynomial is almost $0.5$, which means that it has almost $k/2$ terms.  As can be seen in this table, the PR code with these primitive polynomials can closely approach the Gilbert-Varshamov bound \eqref{eq:GVdisbound} at different rates. It is important to note that one may find other primitive polynomials which can achieve higher minimum Hamming weights at some code rates. Our results show that for sufficiently large $k$ ($k\ge20$) a randomly chosen primitive polynomial of degree $k$ with almost $k/2$ terms can generate PR codes with Hamming weight distribution closely approaching the Binomial distribution and accordingly minimum Hamming weights close to the bound \eqref{eq:mindisbound}, when $n\ge 2k$.

\section{Numerical Results}
In this section, we study the block error rate (BLER) performance of PR codes at fixed block lengths and compare them with eBCH codes with the same dimensions and block lengths. We also provide some results on the rateless performance of PR codes.
\begin{figure}[t]
    \centering
    \includegraphics[width=\columnwidth]{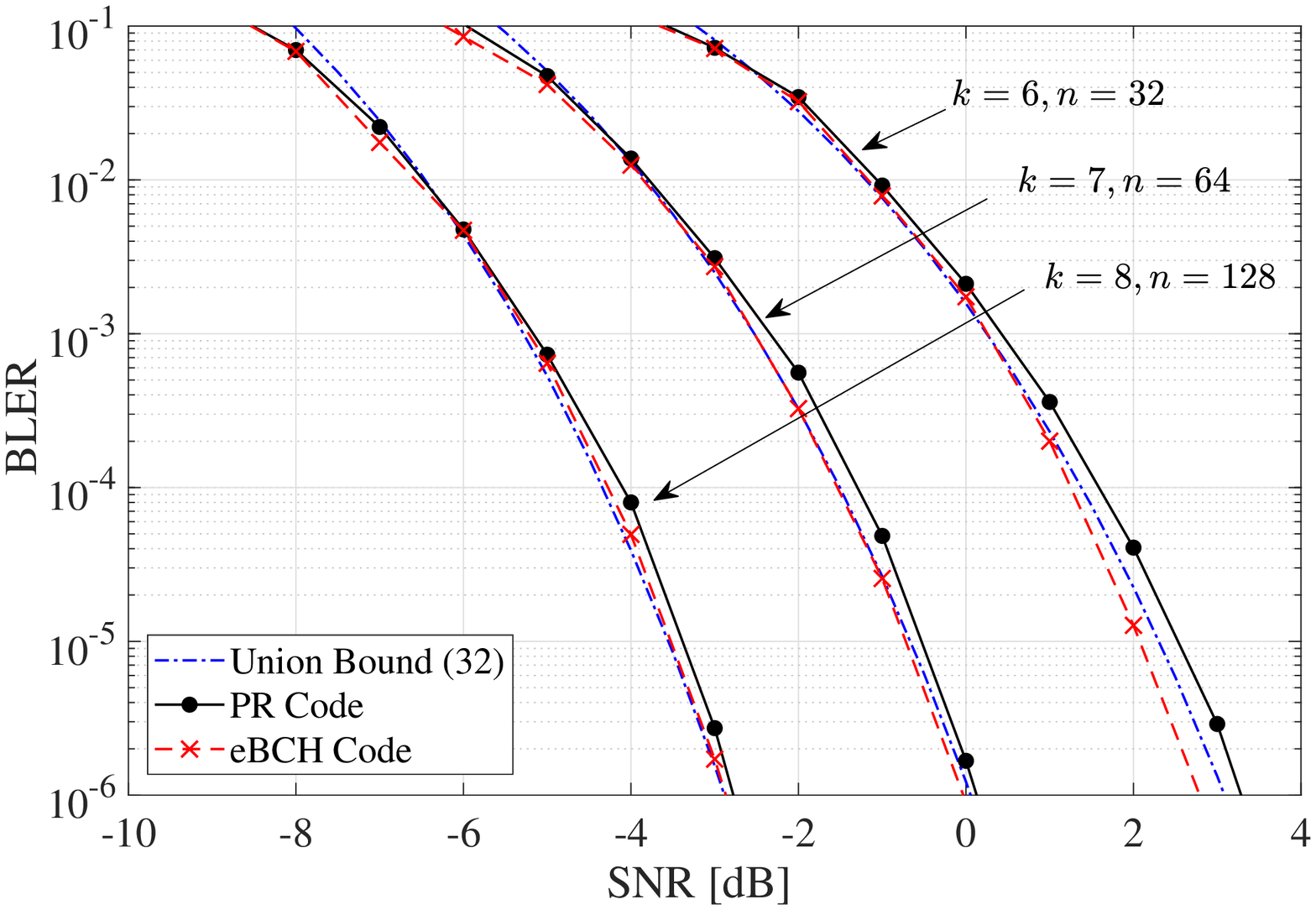}
    \vspace{-2ex}
    \caption{The block error rate performance of the PR and eBCH codes at different block lengths when $k\approx7$. The decoding was performed by using an order-5 OSD.}
    \label{fig:bler7}
\end{figure}
\subsection{Fixed-rate Performance}
A message $\mathbf{b}$ of length $k$ is encoded by using a PR code $\mathrm{PR}(k,n,p(x))$ to generate a codeword $\mathbf{c}$. Each coded symbol $c_i$, for $1\le i\le n$,  is then modulated to $x_i=(-1)^{c_i}$ and sent over a binary-input additive white Gaussian noise (BI-AWGN) channel, $y_i=x_i+w_i$, where $y_i$ is the $i^{th}$ channel output and $w_i$ is the AWGN with variance $N_0/2$. The channel signal-to-noise ratio (SNR) is then given by $\gamma=2/N_0$.

By using the average weight enumerator of PR codes obtained in \eqref{eq:avgWeightPR}, we can derive a union bound (UB) for the BLER as follows:
\begin{align}
    \epsilon_{\mathrm{UB}}=\sum_{i=d_{min}}^{n}\frac{i}{n} \Bar{A}_iQ\left(\sqrt{i\gamma}\right),
    \label{eq:unionbound}
\end{align}
where $Q(.)$ is the standard $Q$-function, $d_{\min}$ is obtained from \eqref{eq:mindisbound}, and $\Bar{A}_i$ is given by \eqref{eq:avgWeightPR}.
\begin{figure}[t]
    \centering
    \includegraphics[width=\columnwidth]{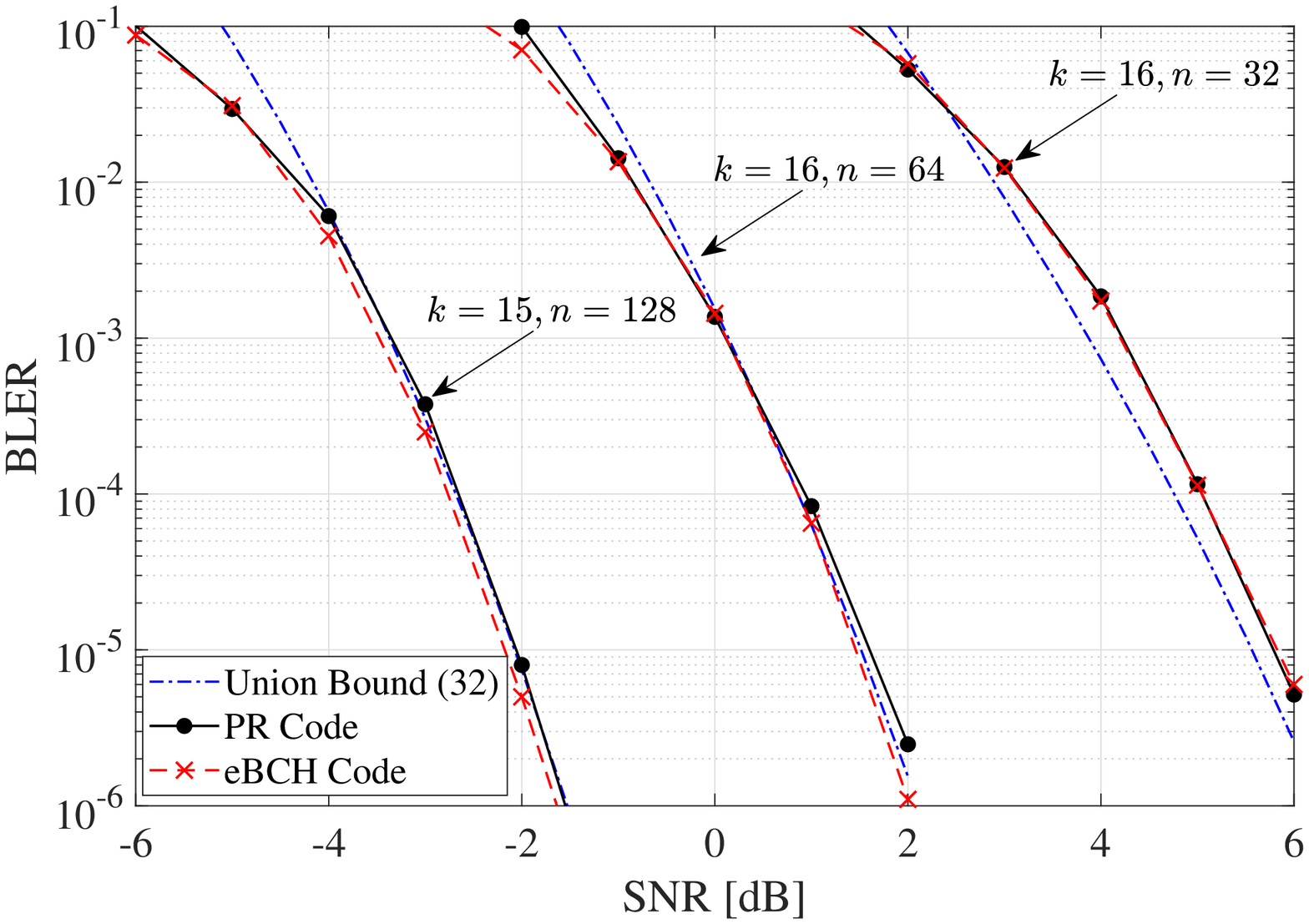}
    \vspace{-2ex}
    \caption{The block error rate performance of the PR and eBCH codes at different block lengths when $k\approx16$. The decoding was performed by using an order-5 OSD.}
    \label{fig:bler16}
\end{figure}
\begin{figure}[t]
    \centering
    \includegraphics[width=\columnwidth]{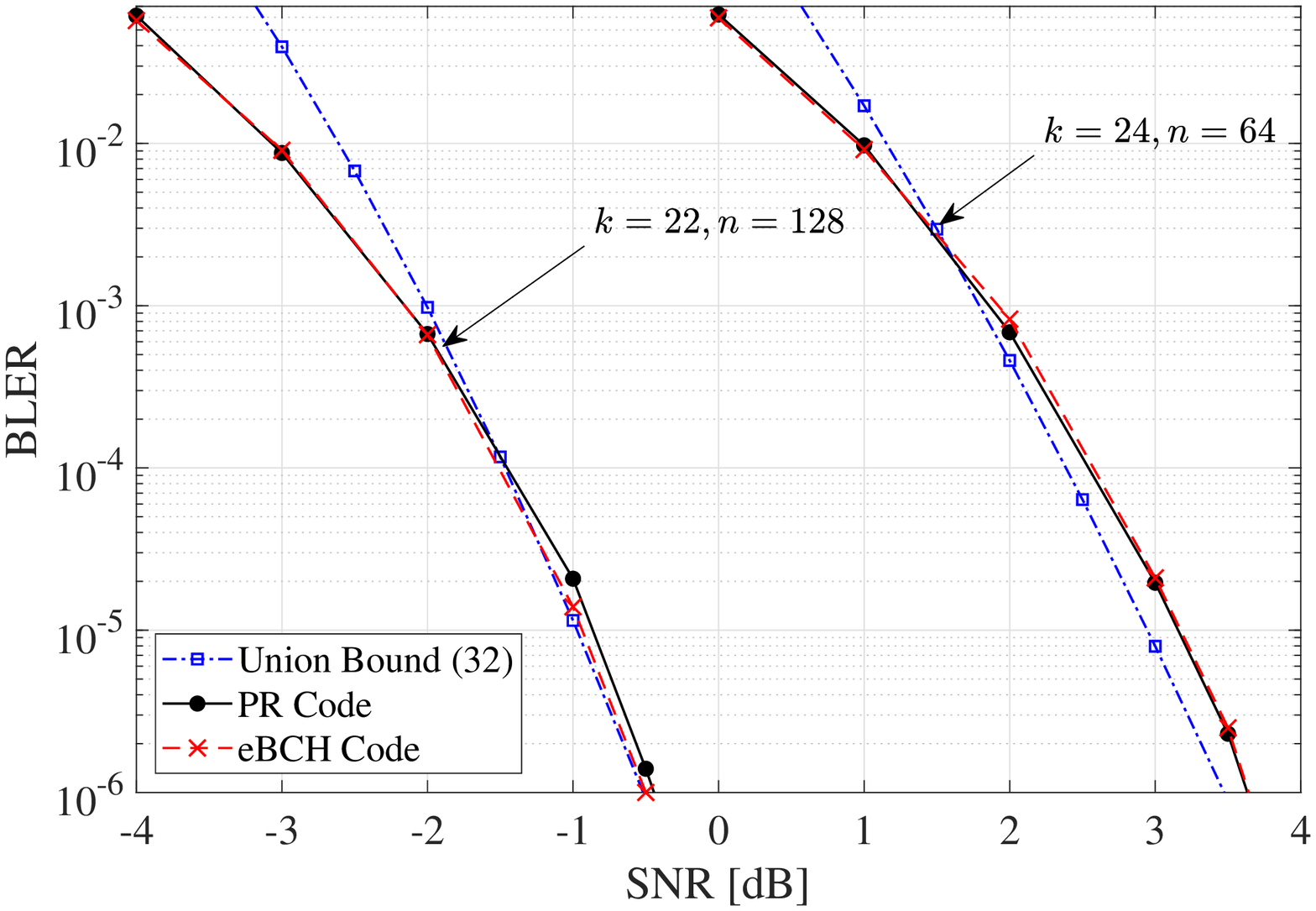}
    \vspace{-2ex}
    \caption{The block error rate performance of the PR and eBCH codes at different block lengths when $k\approx24$. The decoding was performed by using an order-7 OSD.}
    \label{fig:bler24}
\end{figure}
\begin{table*}[t]
\centering
\caption{The weight enumerator of PR codes and eBCH codes. The primitive polynomial for PR codes are obtained from Table II.}
\label{tab3}
\scriptsize
\begin{tabular}{p{0.1cm}p{0.1cm}p{0.6cm}p{14cm}}
     $n$&$k$&Code&Weight Enumerator Polynomial\\
     \hline
     32&6&eBCH&$1+62x^{16} + x^{32}$\\
     &&PR&$1+2x^{12}+4x^{13}+7x^{14}+8x^{15}+14x^{16}+12x^{17}+5x^{18}+8x^{19}+3x^{20}$\\
     \hline
     64&7&eBCH&$1+126x^{32} + x^{64}$\\
     &&PR&$1+2x^{27}+8x^{28}+17x^{29}+23x^{30}+9x^{31}+6x^{32}+7x^{33}+13x^{34}+25x^{35}+13x^{36}+4x^{37}$\\
     \hline
     128&8&eBCH&$1+254x^{64} + x^{128}$\\
     &&PR&$1+3x^{55}+2x^{56}+12x^{57}+7x^{58}+6x^{59}+13x^{60}+14x^{61}+27x^{62}+30x^{63}+17x^{64}+25x^{65}+32x^{66}+\cdots$\\
     \hline
     \hline
     32&16&eBCH&$1+620x^8+13888x^{12}+36518x^{16}+13888x^{20}+620x^{24}+x^{32}$\\
     &&PR& $1+50x^{7}+175x^{8}+455x^{9}+999x^{10}+1953x^{11}+3493x^{12}+5248x^{13}+6944x^{14}+8684x^{15}+9543x^{16}+\cdots$\\
     \hline
     64&16&eBCH&$1+5040x^{24}+12544x^{28}+30366x^{32}+12544x^{36}+5040x^{40}+x^{64}$\\
     &&PR&$ 1+5x^{18}+35x^{19}+79x^{20}+129x^{21}+268x^{22}+525x^{23}+963x^{24}+1485x^{25}+2070x^{26}+2929x^{27}+3889x^{28}+\cdots$\\
     \hline
     128&15&eBCH&$1+8128x^{56}+16510x^{64}+8128x^{72}+x^{128}$\\
     &&PR&$1+x^{46}+7x^{47}+33x^{48}+95x^{49}+135x^{50}+192x^{51}+257x^{52}+397x^{53}+526x^{54}+637x^{55}+935x^{56}+1085x^{57}+\cdots$\\
     \hline
     \hline
     64&24&eBCH&$1+2604x^{16}+10752x^{18}+216576x^{22}+291648x^{24}+1645056x^{26}+888832x^{28}+4419072x^{30}+1828134x^{32}+\cdots$\\
     &&PR&$1+9x^{13}+44x^{14}+112x^{15}+382x^{16}+1180x^{17}+3348x^{18}+8234x^{19}+17863x^{20}+37820x^{21}+73272x^{22}+\cdots$\\
     \hline
     128&22&eBCH&$1+42672x^{48}+877824x^{56}+2353310x^{64}+877824x^{72}+42672x^{80}+x^{128}$\\
     &&PR&$1+2x^{37}+5x^{38}+4x^{39}+12x^{40}+49x^{41}+115x^{42}+275x^{43}+576x^{44}+931x^{45}+1739x^{46}+3155x^{47}+5242x^{48}+\cdots$
\end{tabular}
\end{table*}
Fig. \ref{fig:bler7} shows the BLER performance of PR codes when the message length is $k\approx7$. The primitive polynomial for PR codes are obtained from Table \ref{tab:primpoly}. We use an order-5 ordered statistic decoding (OSD) \cite{Fossorier1995} algorithm for decoding both PR and eBCH codes. As can be seen in this figure, PR codes achieve almost the same BLER as their eBCH counterparts in different SNRs and code rates. The results in Fig. \ref{fig:bler16} and Fig. \ref{fig:bler24} for $k\approx 16$ and $k\approx 24$, respectively, also confirm that PR codes with properly chosen primitive polynomials can achieve BLERs close to their eBCH counterparts. It is important to note that while eBCH codes have relatively larger minimum Hamming weights, PR codes achieves almost the same BLER performance, which is mainly due to the fact the Hamming weight distribution of PR codes is very close to the binomial distribution (when the primitive polynomial is chosen properly), which means that the PR code has a small number of codewords with low Hamming weights. This can be clearly seen in Table \ref{tab3}, that shows the Hamming weight distribution of eBCH and PR codes at different block lengths and rates.

\subsection{Rateless Performance}
We now consider a rateless setting, where we assume that the transmitter wants to deliver a message of length $k$ symbols at the receiver. We assume that the receiver can estimate the channel SNR accurately, however the transmitter does not have any knowledge of the channel SNR. The transmitter uses a PR code $\mathrm{PR}(k,p(x))$ to generate a potentially limitless number of PR coded symbols and continuously send to the receiver. The receiver sends an acknowledgement to the sender when it collected a sufficient number of coded symbols. We use the Polyanskiy-Poor-Verdu (PPV) normal approximation \cite{polyanskiy2010channel} to estimate the number of coded symbols to perform a successful decoding at the desired block error rate. Let $n_s$ denote the number of coded symbols required to perform a successful decoding at the target block error rate $\epsilon_{\mathrm{th}}$. It can be estimated as follows \cite{erseghe2016coding}:
\begin{align}
    n_s=\min_n\left\{n: \epsilon(k,n,\gamma)\le\epsilon_{\mathrm{th}}\right\},
    \label{eq:blppv}
\end{align}
where 
\begin{align}
    \epsilon(k,n,\gamma)=Q\left(\sqrt{\frac{1}{nV(\gamma)}}\left(\frac{nC(\gamma)-k}{\log_2(e)}+\frac{\ln(n)}{2}\right)\right),
    \label{eq:ppvbound}
\end{align}
where for a BI-AWGN channel at SNR $\gamma$, we have
\begin{align}
    C(\gamma)=1+\frac{H^{(1)}(0)}{\ln(2)}, ~~~V(\gamma)=H^{(2)}(0)-\left(H^{(1)}(0)\right)^2,
\end{align}
and 
\begin{align}
    H^{(\ell)}(0)=\frac{1}{\sqrt{2\pi\gamma}}\int_{-\infty}^{-\infty}e^{-\frac{(x-\gamma)^2}{2\gamma}}(-h(x))^\ell dx,
\end{align}
and $h(x)=\ln\left(1+e^{-2x}\right)$ \cite{erseghe2016coding}.

\begin{figure}[t]
    \centering
    \includegraphics[width=\columnwidth]{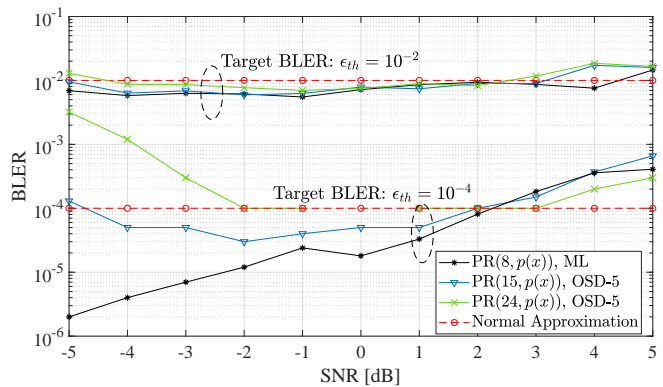}
    \vspace{-2ex}
    \caption{The achievable BLER performance of PR codes at different SNRs for different target BLERs, when $k=8$, $k=16$, and $k=24$. }
    \label{fig:ratelessBLER}
\end{figure}
\begin{figure*}
     \centering
       \subfloat[BI-AWGN channel. \label{fig:ratelessCRC}]{%
       \includegraphics[width=0.49\textwidth]{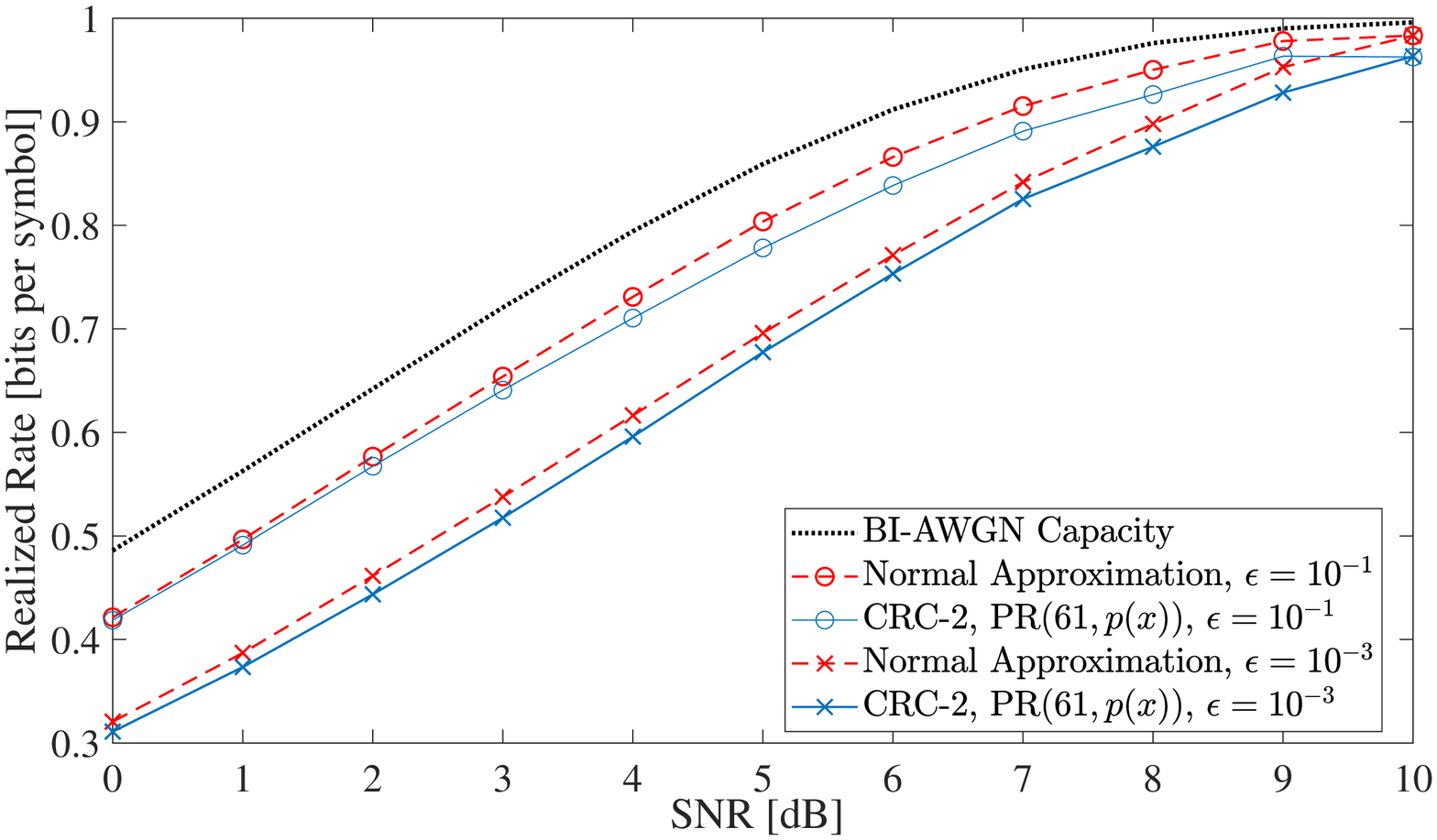}}
    \hfill
    \subfloat[Rayleigh block fading channel with QPSk modulation. \label{fig:ratelessfading}]{%
       \includegraphics[width=0.49\textwidth]{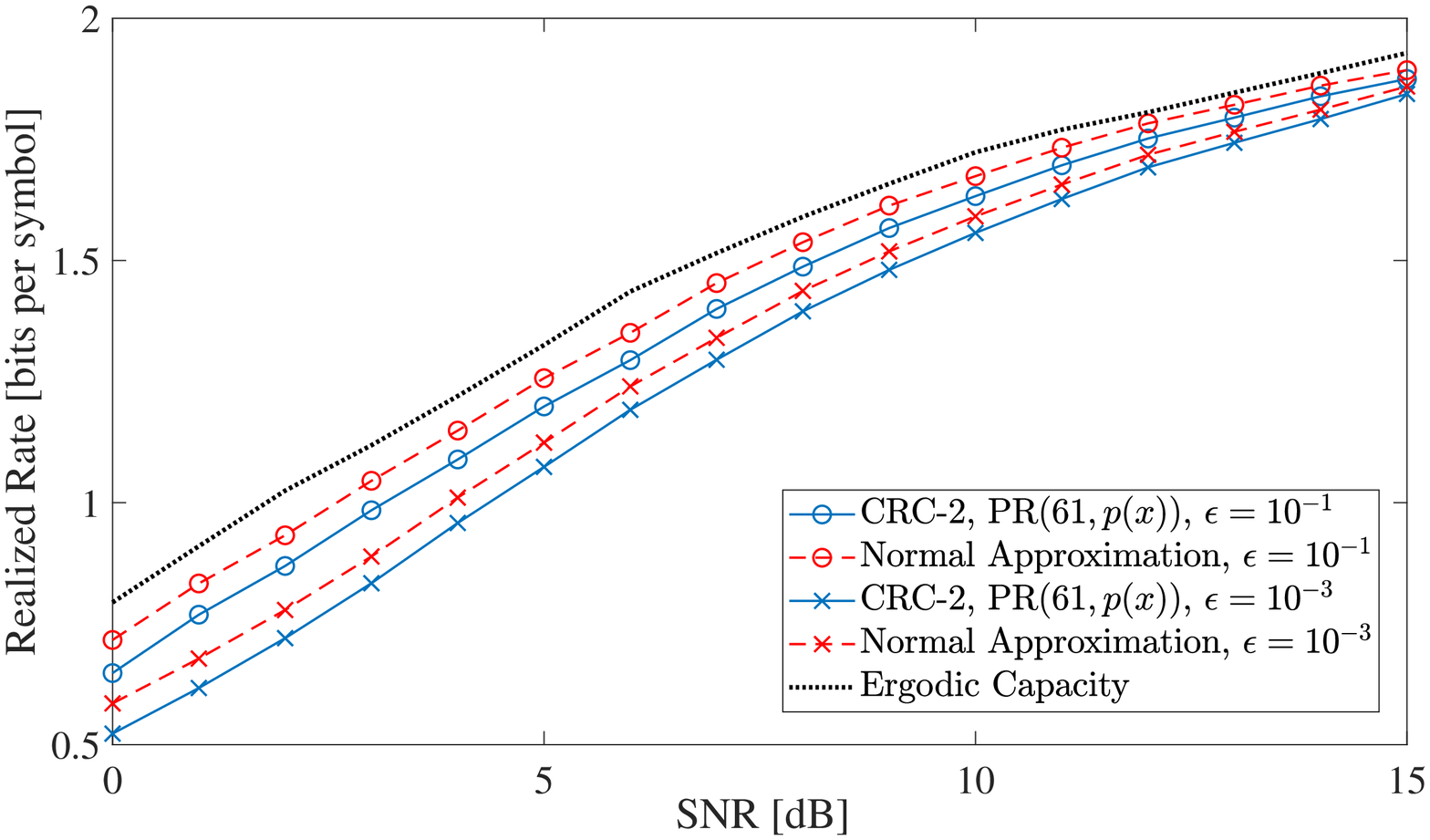}}
    \hfill
        \caption{The achievable throughput of PR code $\mathrm{PR}(61,p(x))$ with 2-bit CRC with the increment of block length $\delta=5$ and initial block length $n_1=n_s$ \eqref{eq:blppv} when an order-4 OSD algorithm was used. $p(x)=1+x+x^{3}+x^{4}+x^{5}+x^{6}+x^{7}+x^{15}+x^{16}+x^{17}+x^{18}+x^{21}+x^{23}+x^{27}+x^{28}+x^{30}+x^{33}+x^{35}+x^{36}+x^{37}+x^{39}+x^{40}+x^{41}+x^{44}+x^{47}+x^{48}+x^{52}+x^{54}+x^{56}+x^{57}+x^{61}$.}
        \label{fig:ratelssfigures}
\end{figure*}

We consider three message lengths, i.e, $k=8,~15$, and $24$, and simulate the PR code in a rateless manner and find the block error rates in various SNRs at block lengths obtained by \eqref{eq:blppv}. Fig. \ref{fig:ratelessBLER} shows the results for the target BLER of $10^{-2}$ and $10^{-4}$ when an order-5 OSD decoder was used for $k=15$ and $k=24$, while the maximum-likelihood (ML) decoder was used for $k=8$. As can be seen the proposed PR code performs very close to the normal approximation bound \eqref{eq:ppvbound} and achieves the target BLER. It is important to note that the approximation (\ref{eq:ppvbound}) looses accuracy at low SNRs, low rates, and very short block lengths; therefore, the estimated number of coded symbols required for a successful decoding may not be accurate. This is the main reason why at high SNRs, there is a large gap between the BLER of PR codes and the target BLER. In particular, when SNR is 5dB, $n_s$ will be very close to $k$, which is very short. The bound in \eqref{eq:ppvbound} is therefore loose and $n_s$ will be inaccurate. It is also important to note that for codes operating at low SNRs, a higher order OSD may be required. This is because for a linear block code $\mathcal{C}(n,k)$ with minimum distance $d_{\mathrm{H}}$ it has been proven that an OSD with the order of $m=\lceil (d_{\mathrm{H}}-1)/4\rceil$ is asymptotically optimum, which means that it can achieve the maximum-likelihood performance. For example, when $\epsilon_{\mathrm{th}}=10^{-4}$, $k=24$, and SNR$=-5$dB, the number of coded symbols obtained by \eqref{eq:blppv} is $n_s=323$ and the respective PR code will have a minimum Hamming weight of $d_{\mathrm{H}}=116$. Therefore, an OSD with an order much larger than 5 is required to have a near optimal performance. The gap at low SNRs for $k=15$ and $k=24$ is mainly due to the low order of the OSD decoder.

We consider another rateless scenario, where the receiver attempts the first decoding when it collected $n_1$ symbols and if successful, it sends an acknowledgment to the transmitter to stop the transmission. We assume that the feedback is instantaneous and error-free. However, if the decoding failed, the receiver collects $\delta>0$ additional symbols and reattempts the decoding using a codeword with $n_1+\delta$ symbols. In particular, in the $i^{th}$ decoding attempt, the receiver has already collected $n_i=n_1+(i-1)\delta$ PR coded symbols and performs the decoding using a codeword of length $n_i$. The transmitter terminates the transmission upon receiving the acknowledgement or when a predetermined number of symbols are sent. We use a $k_c$-bit CRC check to decide whether the decoding in each attempt is successful or not. The throughput or the realize rate of the PR code is then defined as follows:
\begin{align}
    \mathcal{R}=\frac{(k-k_c)(1-\epsilon)}{\mathbb{E}[n]},
\end{align}
where $\mathbb{E}$ is the expectation operand, $n$ is the number of coded symbols collected until the CRC bits are checked, and $\epsilon$ is the block error rate. It is clear that $n$ is random and depends on the noise realization. Fig. \ref{fig:ratelessCRC} shows that the PR code with $k=61$ and 2-bit CRC over the BI-AWGN channel can closely approach the normal approximation bound in a wide range of SNRs at different target BLERs.  We also show the performance of the PR code over the Rayleigh block fading channel with QPSK modulation in Fig. \ref{fig:ratelessfading}. We assumes that the channel state information is available at the receiver, therefore it can determine the initial codeword length $n_1=n_s$ according to \eqref{eq:blppv} to start the decoding. We also assume that the channel remains fixed for the entire duration of decoding a message block of length $k$ bits. As can be seen, the PR code with $k=61$ and 2-bit CRC can closely approach the normal approximation bound in a wide range of SNRs over the fading channel. The performance can be improved by using a higher order OSD algorithm, which in turns increases the complexity.

\subsection{OSD Decoding of PR codes}
Designing an efficient decoding algorithm for PR codes is of critical importance. While this is out of the scope of this paper, we provide some notes on the use of low-complexity OSD algorithms for decoding PR codes.

In OSD, the received coded symbols are first ordered in descending order of their reliability. The generator matrix of the code is accordingly permuted. Next, Gaussian elimination (GE) is performed to obtain the systematic form of the permuted generator matrix. A second permutation may be required during the GE to ensure that the first $k$ columns are linearly independent. The first $k$ bit positions are referred to as the most reliable basis (MRB). Then, MRB will be XORed with a set of test error patterns (TEP) with the Hamming weight up to a certain degree, referred to as the order of OSD. Then the vectors obtained by XORing the MRB are re-encoded using the permuted generator matrix to generate candidate codeword estimates. The codeword estimate with the minimum distance from the received signal is selected as the decoding output.

OSD is an approximate maximum likelihood (ML) decoder for block codes \cite{Fossorier1995}. More specifically, for a linear block code $(n, k)$ with minimum Hamming weight $d_{\min}$, it is proven that an OSD with order $m = \lceil d_{\min}/4 - 1\rceil$ is asymptotically optimum \cite{Fossorier1995}. OSD is, however, complex and its algorithmic complexity can be up to $\mathcal{O}(k^m)$ for an order-$m$ OSD. Several approaches have been recently proposed to reduce the number of TEPs required to be re-encoded to find the best codeword estimate. Authors in \cite{Chentao2021OSD} characterized the evolution of the distance distribution during the reprocessing stage of the OSD algorithm. They accordingly proposed several decoding rules, namely sufficient and necessary conditions, to reduce the complexity significantly. These are mainly to terminate the decoding early, when a suitable candidate codeword is found, and to discard TEPs, which are less likely to generate promising codeword candidates.  Other approaches introduced in \cite{Chentao2019SDOSD,Chentao2020PBOSD} can also be used to further reduce the complexity by searching through the TEPs in an optimal manner, which will result in finding the best codeword estimate faster. An efficient implementation in C has shown that the OSD decoding with sufficient and necessary conditions can run an order-5 OSD in a few $\mu$s per codeword \cite{Choi2019FastOSD}.

For the simulations in this paper, we used the simple probabilistic necessary condition (PNC) proposed in \cite{Choi2019FastOSD} to terminate the decoding when a candidate codeword is found with the distance to the received signal lower than a certain threshold. The threshold value for the $i^{th}$ reprocessing order is calculated as $S_i = \sum_{k-1-i}^{k-1}|\bar{y}_i| + \beta(n - k)$, where $\bar{y}_i$ is the re-ordered received signal. In the $i^{th}$ reprocessing stage, once a codeword with distance to the received codeword less than $S_i$ is found, the decoder terminates and skips the remaining orders. For example, when decoding the PR code with $k=22$ and $n=128$ (Fig. \ref{fig:bler24}) using an order-7 OSD with PNC \cite{Choi2019FastOSD} and $\beta=0.08$, we only need to check on average 1265 TEPs at SNR$=-1$dB. This is a significant reduction from 280599 TEPs in the original order-7 OSD, while achieving the same BLER performance. The decoding run-time per codeword is accordingly reduced by two orders of magnitude. Further reductions in the number of TEPs and running time can be achieved by using approaches proposed in \cite{Chentao2021OSD}.

We note that other decoding approaches, such as the Berlekamp–Massey algorithm \cite{massey1969shift}, can be modified to decode PR codes. This is however out of the scope of this work and will be discussed in future works.

\section{Conclusions and Future Works}
In this paper, primitive rateless (PR) codes were proposed. A PR code is mainly characterized by the message length $k$ and a primitive polynomial of degree $k$, where the $i^{th}$ columns of the generator matrix is the binary representation of $\alpha^{i-1}$, where $\alpha$ is a primitive element of $\mathbf{GF}(2^k)$ and is the root of $p(x)$. We showed that a PR code can be also constructed 1) by using a linear-feedback shift-register (LFSR) with connection polynomial $x^kp(1/x)$ and 2) by using Boolean functions. We proved that any two PR codes of dimension $k$ and truncated at length $n\ge 2k$, which are constructed by using two distinct primitive polynomials, do not have any non-zero codeword in common. We characterized the average Hamming weight distribution of PR codes and developed a lower bound on the minimum Hamming weight which is very close to the Gilbert-Varshamov bound. We  proved that for any $k$, there exists at least one PR code that can meet this bound. We further found some good primitive polynomials for PR codes of dimension $k\le 40$ which can closely approach the Gilbert-Varshamov bound. Simulation results show that the PR code with a properly chosen primitive polynomial can achieve similar block error rate performance as the eBCH code counterpart. We further simulated the PR code in a rateless setting and showed that it can achieve very high realized rates over a wide range of SNRs. PR codes can be designed for any message length and the primitive polynomial can be optimized for any block length. Potential future directions could be finding a framework to optimized the primitive polynomial and devising novel on-the-fly decoding approaches for PR codes. 

\bibliographystyle{IEEEtran}
\footnotesize
\bibliography{ref}

\begin{IEEEbiography}[{\includegraphics[width=1in,height=1.25in,clip,keepaspectratio]{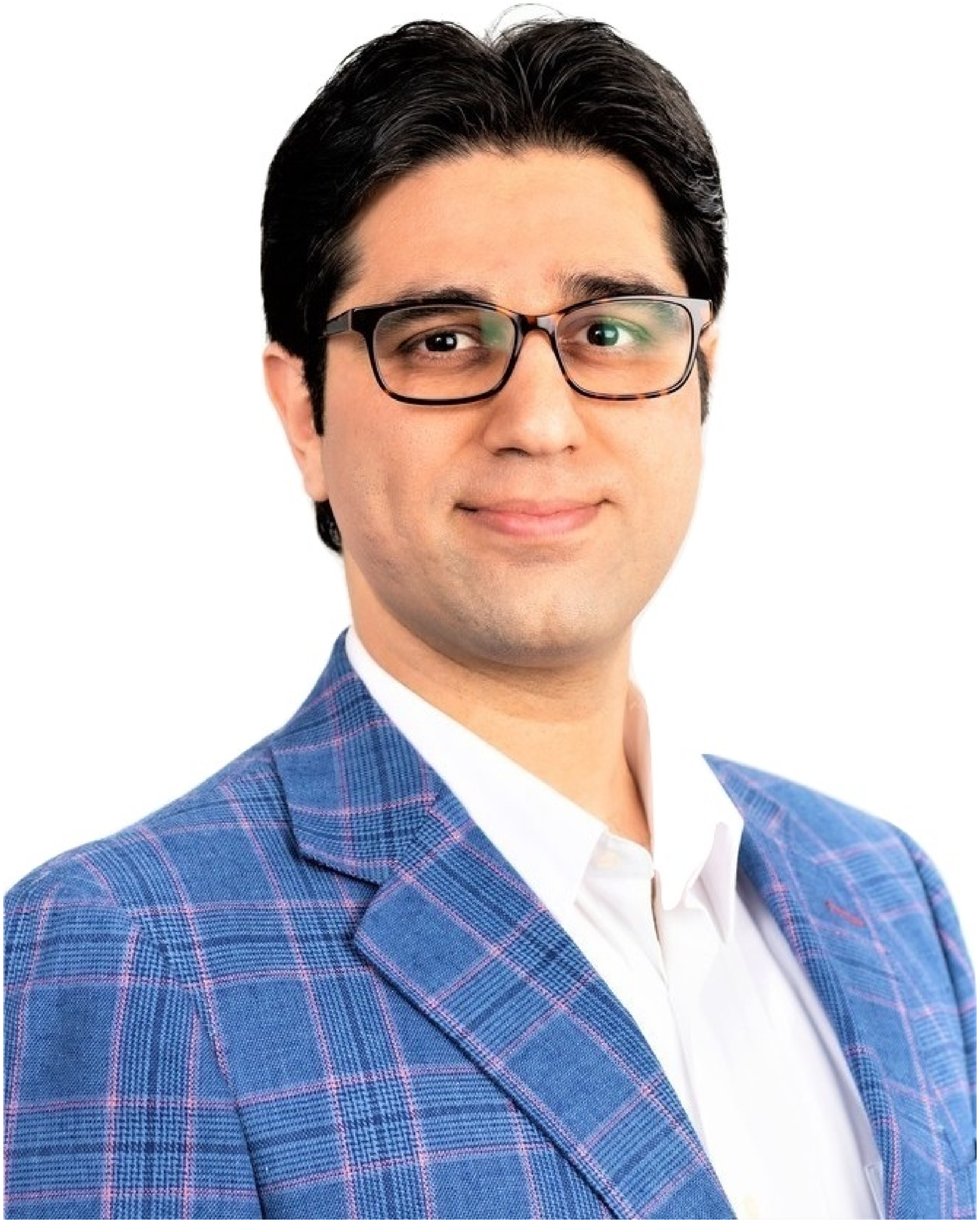}}]{Mahyar Shirvanimoghaddam} 
 (Senior Member, IEEE) received the B.Sc.
degree (Hons.) from The University of Tehran, Iran, in 2008, the M.Sc. degree (Hons.) from Sharif University of Technology, Iran, in 2010, and the Ph.D. degree from The University of Sydney, Australia, in 2015, all in Electrical Engineering. He is currently a Lecturer with the Centre for IoT and Telecommunications, The University of Sydney. His research interests include coding and information theory, rateless coding, communication strategies for the Internet of Things, and information-theoretic approaches to machine learning. He is a fellow of the Higher Education Academy. He was selected as one of the Top 50 Young Scientists in the World by the World Economic Forum in 2018 for his contribution to the Fourth Industrial Revolution. He received the Best Paper Award for the 2017 IEEE PIMRC, The University of Sydney
Postgraduate Award and the Norman I Prize, and The 2020 Australian Award for University Teaching. He also
serves as a Guest Editor for the \emph{Journal of Entropy} and \emph{Transactions on
Emerging Telecommunications Technologies}.
\end{IEEEbiography}

\end{document}